\documentclass[5p]{elsarticle} 

\usepackage{lineno}
\modulolinenumbers[5]

\usepackage{mathtools}
\journal{Energy Conversion and Management}

\usepackage{placeins}

\usepackage{eurosym}
\usepackage{threeparttable} 

\usepackage{url}
\usepackage[colorlinks=true, citecolor=blue, linkcolor=blue, filecolor=blue,urlcolor=blue]{hyperref}

\usepackage{lineno,hyperref}
\modulolinenumbers[1]
\usepackage{amsmath}
\usepackage{siunitx}
\usepackage{eurosym}
\biboptions{numbers,sort&compress}
\usepackage[europeanresistors,americaninductors]{circuitikz}
\usepackage{adjustbox}
\usepackage{xspace}
\usepackage{caption}
\usepackage{booktabs}
\usepackage{tabularx}
\usepackage{threeparttable}
\usepackage{multicol}
\usepackage{float}
\usepackage{graphicx,dblfloatfix}
\usepackage{csvsimple}
\newcommand{\ubar}[1]{\text{\b{$#1$}}}
\newcommand*\OK{\ding{51}}

\def\el{${}_{\textrm{el}}$}
\def\th{${}_{\textrm{th}}$}

\begin{document}

\begin{frontmatter}

\title{The role of storage technologies throughout the decarbonisation of the sector-coupled European energy system}

\author[mymainaddress,iClimate]{Marta Victoria\corref{mycorrespondingauthor}}
\ead{mvp@eng.au.dk}
\author[mymainaddress]{Kun Zhu}
\author[kitaddress]{Tom Brown}
\author[mymainaddress,iClimate]{Gorm B. Andresen}
\author[mymainaddress,iClimate]{Martin Greiner}
\cortext[mycorrespondingauthor]{Corresponding author}
\address[mymainaddress]{Department of Engineering, Aarhus University, Inge Lehmanns Gade 10, 8000 Aarhus, Denmark}
\address[iClimate]{iCLIMATE Interdisciplinary Centre for Climate Change, Aarhus University}
\address[kitaddress]{Institute for Automation and Applied Informatics (IAI), Karlsruhe Institute of Technology (KIT), Forschungszentrum 449, 76344, Eggenstein-Leopoldshafen, Germany}

\begin{abstract}

We use an open, hourly-resolved, networked model of the European energy system to investigate the storage requirements under decreasing CO$_2$ emissions targets and several sector-coupling scenarios. For the power system, significant storage capacities only emerge for CO$_2$ reductions higher than 80\% of 1990 level in that sector. For 95\% CO$_2$ reductions, the optimal system includes electric batteries and hydrogen storage energy capacities equivalent to 1.4 and 19.4 times the average hourly electricity demand. Coupling heating and transport sectors enables deeper global CO$_2$ reductions before the required storage capacities become significant, which highlights the importance of sector coupling strategies in the transition to low carbon energy systems. A binary selection of storage technologies is consistently found, \textit{i.e.}, electric batteries act as short-term storage to counterbalance solar photovoltaic generation while hydrogen storage smooths wind fluctuations. Flexibility from the electric vehicle batteries provided by coupling the transport sector avoid the need for additional stationary batteries and reduce the usage of pumped hydro storage. Coupling the heating sector brings to the system large capacities of thermal energy storage to compensate for the significant seasonal variation in heating demand.

\end{abstract}


\end{frontmatter}


\section{Introduction}
\label{sec_introduction}

The IPCC \textit{Special Report on Global Warming of 1.5$^{\circ}$C} \cite{IPCC_1.5} has shown that the European Union's commitments to decrease CO$_2$ emissions, which include a reduction target of 80-95\% by 2050 relative to 1990 levels, are not sufficient. Limiting global warming to 1.5$^{\circ}$C compared to 2$^{\circ}$C or more reduces the risks associated with long-lasting and irreversible changes. To that end, the decarbonisation of our economy must be deep and fast, and net-zero carbon emissions need to be achieved globally by 2050. 
In that context, the European Commission has recently called for a climate-neutral Europe by 2050 \cite{in-depth_2018}. One of the prominent strategies to supply Europe's final energy consumption with very low CO$_2$ emissions relies on the installation of vast capacities of Variable Renewable Energy Sources (VRES), \textit{i.e.}, wind and solar photovoltaics. The low generation costs achieved by these technologies allow us to envisage a future in which VRES supply a significant share not only of the electricity demand but of the final energy consumption. The fundamental challenge of this strategy is how to counterbalance the fluctuating generation as we approach high renewable penetration.  One of the first answers that comes to mind is storage. \\

Using weather-driven renewable energy system modelling with hourly resolution and assuming a simple storage dispatch algorithm, 
several authors have identified a divergence of the required storage energy capacity when the average renewable generation approaches the average electricity demand \cite{Heide_2011, Rasmussen_2012, Jensen_2014}. In their models, costs are not included and the required storage energy and power capacities are determined by the most critical situations throughout the year. Moreover, the weather-driven modelling has also been used to investigate the link between the storage energy capacity and the favoured VRES generation, either wind or solar, and vice versa. For instance, storage energy capacities in the range of 6 times the average hourly demand are sufficient to drastically reduce the number of hours in which VRES generation is not enough to supply demand and, hence, minimise the required backup generation \cite{Rasmussen_2012}. The reason behind is that the pronounced diurnal pattern of solar generation can be smoothed by such stores which charge during the day and discharge throughout the night. Similarly, the provision of additional short-term storage by allowing smart charging and discharging into the grid of a future European fleet of Electric Vehicles (EVs) could ease the integration of large shares of solar electricity. Hydrogen storage with energy capacity in the range of several days of average consumption were found to be adequate to counterbalance wind fluctuations in the synoptic time scale \cite{Rasmussen_2012}. \\

Andresen and coauthors \cite{Andresen_2014} applied the weather-driven modelling to Denmark and investigated the required energy and power capacity of storage to minimise VRES surplus. They found that, for regions where wind is a better resource, only large-scale seasonal storage, \textit{e.g.} hydrogen tanks, enables the use of VRES surplus to cover electricity demand at any time. Bussar \textit{et al.} analysed the possibility of fully decarbonising the electricity generation in the EUMENA (Europe, Middle East, and North Africa regions) using a rule-based dispatch of system components and calculating the required storage capacity through genetic optimisation \cite{Bussar_2014, Bussar_2016}. Weather-driven and rule-based models are very useful to understand the general dynamics of energy systems, but, by neglecting the costs, they can overestimate storage requirements. Besides, it is not obvious how to set up the dispatching rules when the power system is coupled with other sectors, as it will be introduced later. \\

Schlachtberger and coauthors followed a different approach in which they minimised the annual costs of the European power system represented by a network of interconnected countries \cite{Schlachtberger_2017}. The optimal system configuration includes larger solar capacities in southern countries accompanied by electric batteries whose discharge time at maximum power was assumed to be 6 hours. Countries in north-western Europe, with better onshore and offshore wind resources, make use of these technologies together with long-term hydrogen storage capacities (1-week discharge time is assumed in \cite{Schlachtberger_2017}) and an extension of the grid infrastructure to allow a synoptic-scale temporal and spatial smoothing of wind fluctuations. Cebulla \textit{et al.} \cite{Cebulla_2017} also employed cost optimisation to investigate the required storage capacities in the European power system represented by a network of 7 regions with 89\% renewable share. They found that lithium-ion batteries are selected in regions with high solar generation while hydrogen storage and adiabatic Compressed Air Energy Storage (aCAES) is deployed in regions with high wind generation. In \cite{Cebulla_2018}, the electricity storage optimal power and energy capacities calculated by different authors are plotted together and their dependence with the renewable penetration and the solar and wind mix are investigated. 
\\

Storage represents one possible strategy to cope with VRES intermittent generation but there are at least two alternatives. The first one consists in extending the transmission capacities among adjacent countries. Detailed analysis of the potential of this strategy to enable a highly renewable European power system can also be found in references \cite{Heide_2011,Schlachtberger_2017,Rodriguez_2014,Eriksen_2017,Gils_2017a}. The second alternative relies on coupling the power system with other sectors such as heating or transport. The non-electric sectors not only provide additional flexibility to integrate VRES but also get access to low-carbon energy generation that reduces the CO$_2$ emissions from these sectors. Estimations on how to provide electricity and heating demand using wind, solar energy and storage were already presented in 1975 by S{\o}rensen \cite{Sorensen_1975}, but his calculations were restricted to Denmark. A scenario-based analysis of the potential of sector-coupling to decarbonise the European energy system was published by Connolly \textit{et al.} who named this approach `Smart Energy System' and assumed infinite interconnection capacity among European countries \cite{Connolly_2016}. \\

Brown and coauthors \cite{Brown_2018} introduced the first open-source, sector-coupled, networked model for the European energy system that we used here. The authors performed a global analysis of the synergies of sector coupling in which the benefits of the successive integration of different sectors are investigated together with the impacts of extending transmission capacities among European countries. In addition, the evolution of the interactions among the sectors as CO$_2$ emissions approach zero was analysed in \cite{Brown_2019}. We have also investigated the coupling of the heating and power sectors under variable CO$_2$ price \cite{Zhu_2019}. Child and coauthors  \cite{Child_2018, Child_2019} performed a brownfield cost optimisation every 5 years of the European energy system to model the transition up to 2050. Besides centralised storage technologies (batteries, PHS, aCAES, gas, and thermal energy storage), their model includes home-based batteries associated with rooftop solar PV systems whose global energy capacity is independently optimised by a separated modelling of prosumers behaviour. \\

In this work, a networked sector-coupled model of the European energy system is cost-optimised under decreasing CO$_2$ emissions targets to investigate the following research questions:\\

1. What are the required storage capacities as CO$_2$ emissions curb? Moreover, we want to evaluate the sensitivity of the cost-effective storage capacities to the assumptions and constraints imposed on the model. \\

2. What is the impact of sector-coupling in the storage requirements?  We want to analyse the consequences of adding storage technologies provided by sector-coupling, \textit{e.g.} thermal energy storage in the heating sector or EV batteries in the transport sector, and what are the characteristic operating frequencies of the different technologies. \\

The most novel contribution of the results presented here is the systematic study of the storage requirements as a function of the CO$_2$ reduction target and the coupling of the European power system with the heating and transport sectors. Other studies have quantified the required storage for a European power system with high renewable penetration \cite{Rasmussen_2012, Jensen_2014, Schlachtberger_2017,Eriksen_2017, Gils_2017a, Bussar_2014, Cebulla_2017, Cebulla_2018}. This study extends the existing body of literature by analysing the storage requirements under CO$_2$-emissions-capped cost-optimisation and investigates the dispatch pattern of different storage technologies and how they influence each other.\\

The paper is organised as follows. Section \ref{sec_storage_technologies} introduces the storage technologies considered in the analysis. Section \ref{sec_methods} briefly introduces the methodology which is thoroughly described in \ref{annex_pypsa_model} and \ref{anex_data}, where details on the model and data are provided. Section \ref{sec_results} gathers the primary outcomes of this work and compares them to previous results in the literature. Section \ref{sec_conclusions} summarises the main conclusions.  

\section{Storage technologies} \label{sec_storage_technologies}

Table \ref{tab_stores_characteristics} shows the costs, lifetime, and efficiency values assumed for the storage technologies investigated in this paper. In all the scenarios, described in Section \ref{sec_scenarios}, electricity can be stored in electric batteries, overground hydrogen storage, and pumped hydro storage (PHS). Sector coupling provides additional storage in two different ways. On the one hand, in scenarios which include the transport sector, electricity can also be stored in EV batteries. On the other hand, the heating sector includes individual short-term (ITES) and centralised long-term thermal energy storage (CTES).

\begin{figure}[ht!]
\centering
\includegraphics[width=0.7\columnwidth]{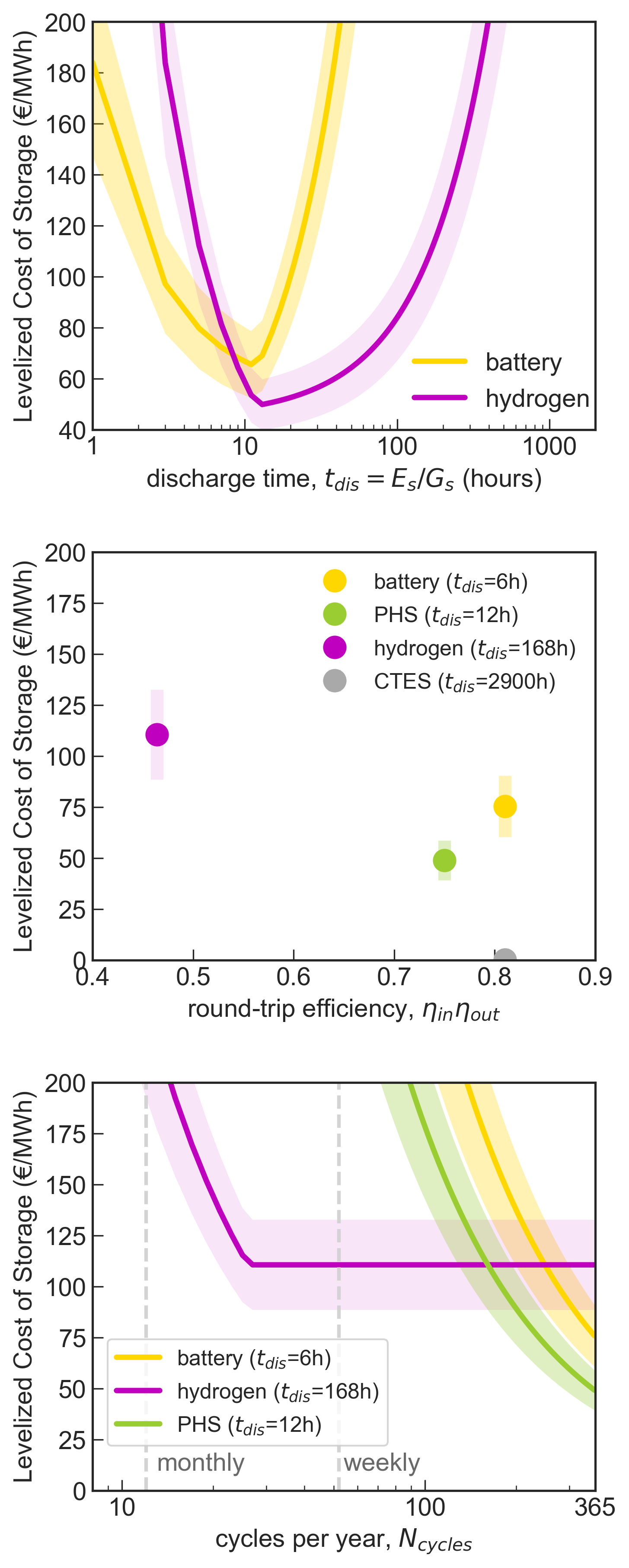}\\
\caption{ Simplified analysis without considering the rest of the system. (top) LCOS of as a function of the discharge time $t_{dis}$ (ratio between the storage energy capacity $E_s$ and power capacity $G_s$), (middle) LCOS vs round-trip efficiency of different storage technologies, (bottom) LCOS vs number of cycles per year. In the middle and bottom figures, the following discharge times have been assumed: battery 6 h, PHS 12 h, hydrogen 1 week, CTES 4 months. When calculating $N_{cycles}$, the following limits are assumed: PHS and batteries can cycle daily, and hydrogen storage is operating at maximum power every hour, either charging or discharging. The shadow areas represent LCOS when costs assumed for the different technologies (see Table \ref{tab_stores_characteristics}) vary $\pm$ 20\%. Null price for energy input to the stores is assumed to estimate LCOS, Eq. (\ref{eq_LCOS_storage}).} \label{fig_storage} 
\end{figure}

\begin{table*}
\begin{threeparttable}
	\centering
		\caption{Characteristics of storage technologies. For batteries, hydrogen, and thermal energy storage, power and energy capacities are independently optimised.} \label{tab_stores_characteristics}
		\
		\begin{footnotesize}

    \begin{tabular}{p{5.5cm}p{1.1cm}p{1.1cm}p{2.2cm}p{2.0cm}p{0.9cm}p{1cm}p{1.1cm}}
		\hline
     & Energy  capacity (GWh) & Power capacity (GW) & Round-trip efficiency $\eta_{in} \cdot \eta_{out}$ & Overnight Cost & FOM\tnote{a} (\%/a) & Lifetime (years) & Source\\
		\hline		
    Pumped hydro storage (PHS)\tnote{b}  & 285 & 47.5 & 0.87$\cdot$0.87=0.76 & 2000 \EUR/KW$_e$ & 1     & 80 & \cite{Schroeder_2013}\\
    Battery & optimised &  & 0.9$\cdot$0.9=0.81 & 144.6 \EUR/KWh$_e$ & 0     & 15 & \cite{Budischak_2013}\\
    Battery inverter &  &  optimised & 0.9 & 310 \EUR/KW$_e$ & 3     & 20 & \cite{Budischak_2013} \\
    Hydrogen storage & optimised &  & 0.8$\cdot$0.58=0.46 & 8.4 \EUR/KWh$_e$ & 0     & 20 & \cite{Budischak_2013, Steward_2009}\\
    Hydrogen electrolysis &  &  optimised & 0.8 & 350 \EUR/KW$_e$ & 4     & 18 & \\
    Hydrogen fuel cell &  & optimised & 0.58 & 339 \EUR/KW$_e$ & 3     & 20 & \cite{Budischak_2013, Steward_2009}\\
    EV batteries\tnote{c} & 6,150 & 1,350 & 0.9$\cdot$0.9=0.81 &  &       &  & \\
    Individual thermal energy storage (ITES) & optimised & optimised & 0.9$\cdot$0.9=0.81 & 860 \EUR/m$^3$ & 1     & 20 &  \cite{Henning_2014, Lund_2016}\\
    Long-term thermal energy storage (CTES) & optimised & optimised & 0.9$\cdot$0.9=0.81 & 30 \EUR/m$^3$ & 1     & 40 & \cite{Henning_2014, Lund_2016}\\
		\hline
  
		\hline

		\end{tabular}
		\end{footnotesize}
		\begin{tablenotes}
		\item [a] \footnotesize{Fixed Operation and Maintenance (FOM) costs are given as a percentage of the overnight cost per year.}
    \item [b] \footnotesize{Pumped Hydro Storage (PHS) is exogenous to the system. The capacity in every country is fixed and considered to be fully amortised. }
		\item [c] \footnotesize{EV batteries are exogenous to the system; their cost is not included in the optimisation.}

\end{tablenotes}
\end{threeparttable}
\end{table*}

\paragraph{Electric batteries}\ 

Cost predictions for batteries have significant uncertainties. A greater demand in the coming years might increase the annual manufacturing rate and the learning curve might be travelled faster than expected, decreasing costs \cite{Schmidt_2017}. Moreover, increased R\&D investments could also accelerate the achievement of cost-effective electricity storage \cite{Kittner_2017}. We follow \cite{Budischak_2013} for the cost assumptions of batteries and hydrogen storage but acknowledge the uncertainty associated with these technologies and evaluate the impacts of the cost assumptions by the sensitivity analysis included in Section \ref{sec_sensitivity}. For central electricity storage, lithium-titanate batteries are considered with a lifetime of 15 years and 144.6 \EUR /kWh  \cite{Budischak_2013}. Lithium-titanate is a type of lithium-ion battery preferred because it has the most extended cycle life and the ability to charge quickly. The DC/AC converter is assumed to have an efficiency of 90\% for every conversion, and a cost of 310 \EUR /kW$_e$ \cite{Budischak_2013}. Recycling electric batteries is challenging \cite{recycle_batteries} yet necessary to ensure the sustainability of future energy systems. Due to the lack of available information, the cost of recycling is not included here, but its effect could be evaluated in light of the sensitivity analysis.\\

\paragraph{Hydrogen storage}\ 

Electricity is converted into hydrogen through electrolysers with 80\% conversion efficiency and at a cost of 350 \EUR /kW$_e$ \cite{Budischak_2013}. Fuel cells are used to convert hydrogen back into electricity with 58\% conversion efficiency and 339 \EUR/kW$_e$ \cite{Budischak_2013, Steward_2009}. Hydrogen can be stored underground in salt caverns or overground in steel tanks \cite{Steward_2009}. The second option has been considered here which avoids constraining the maximum energy capacity in every scenario by the availability of the suitable geological formations in every country. The cost assumed for hydrogen storage in overground steel tanks is 8.4 \EUR /kWh \cite{Budischak_2013, Steward_2009}. However, it is important to remind that the existing energy capacity in salt cavern in Europe is huge, see for instance \cite{Staffell_2019, Gilhaus_2007, Cebulla_2017}, and such installations at a lower scale are already in operation \cite{sunstorage, Staffell_2019}.\\

\paragraph{Preliminary analysis: Electric batteries vs Hydrogen storage}\

The Levelized Cost of Storage (LCOS) is defined as the total lifetime cost of the investment divided by its cumulative delivered energy \cite{Julch_2016, Schmidt_2019}. For a certain storage technology $s$, LCOS can be estimated as 

\begin{equation} \label{eq_LCOS_storage}
LCOS = \frac{\frac{c_{n,s}}{t_{dis}} + \hat{c}_{n,s} + \sum_{t} p_{t}g_{s,t}}{\eta_{in} \eta_{out}\cdot N_{cycles}}
\end{equation}
where $c_{n,s}$ and $\hat{c}_{n,s}$ are the annualised costs of the power and energy capacity (in \EUR /kW and \EUR /kWh) which include investment cost and operation and maintenance (see Table \ref{tab_costs}), $t_{dis}$ is the discharge time (in hours) at maximum power, $\eta_{in}\eta_{out}$ is the round-trip efficiency of the storage, and $N_{cycles}$ the number of cycles in a year \cite{Pawel_2014, Julch_2016, Schmidt_2019}. End-of-life costs are not included in this analysis. For the preliminary analysis in Figure \ref{fig_storage}, the price $p_{t}$ of the energy $g_{s,t}$ input in the store is assumed to be zero. A discount rate of 0.07 is used to calculate annualised costs. Figure \ref{fig_storage} depicts the LCOS as a function of the discharge time, the round-trip efficiency, and the number of cycles in a year, for different storage technologies. To provide an initial comparison, the discharge time for every technology is fixed  in the middle and bottom plots of Figure \ref{fig_storage}, but in the following analyses (Section \ref{sec_model} and following), power and energy capacities are independently optimised. \\

On the one hand, the unitary energy capacity cost for hydrogen storage is significantly lower than that of batteries, but on the other, the unitary power capacity cost is higher for hydrogen storage and the round-trip conversion efficiency is lower. As shown in the top plot of Figure \ref{fig_storage}, based on preliminary analysis, batteries are cost-effective for short-term storage, while, under these cost assumptions, for $t_{dis}>$ 9 hours, it is cheaper to use hydrogen storage. The plot on the bottom of Figure \ref{fig_storage} depicts the LCOS as a function of the number of cycles in a year. For daily cycling, electric batteries are cost-competitive but their cost dramatically increases for lower dispatching frequencies where hydrogen storage becomes a better option. So far, a straightforward model, Eq.  (\ref{eq_LCOS_storage}) has been used to estimate the LCOS, which assumes no cost for the energy input in the store. J{\"u}lch \textit{et al.} \cite{Julch_2016} and Schimdt \textit{et al.} \cite{Schmidt_2019} applied similar simple models to determine the most cost-effective storage technology for different system applications. Both authors assumed a constant price for the electricity input in the stores. However, in reality, storage technologies store energy when the market price is low and put it back into the system when the market price is high. By doing so, they facilitate the system operation and reduce the total system cost. Therefore, to calculate the capacities of the different storage technologies that result cost effective, the system components, \textit{i.e.}, generators, stores, interconnection, etc. can not be independently determined but their capacity and dispatch time series must be jointly optimised. This is the strategy that we follow in this paper and which is described in Section \ref{sec_model}.

\paragraph{Pumped Hydro Storage (PHS)} \

PHS represents the most abundant electricity storage in Europe with 47.5 GW installed capacity. However, the potential expansion of this technology is low as most of the convenient locations are already under exploitation. Hence, we consider PHS exogenous to the model, that is, it is not subject to optimisation but fixed at its current capacity for every country \cite{Schroeder_2013}. The round-trip efficiency considered for PHS is 76\%. Besides PHS, run-of-river and reservoir hydropower plants are included in the model, and they are described in \ref{sec_electricity}.

\paragraph{Electric Vehicle (EV) batteries} \

The potential energy storage capacity provided by EV batteries is estimated by assuming that the cars fleet in Europe (246 million vehicles, 0.465 per population of 529 million people) is transformed into battery electric vehicles (BEV) with an average energy storage capacity of 50 kWh and charging capacity of 11 kW. The charging efficiency is assumed to be 90\%. The BEVs are modelled in aggregate in every country, it is considered that half of them can shift their charging time as well as discharge into the grid to facilitate the operation of the system and reduce its total cost. Hence, the total available EV batteries energy storage capacity contributing to the power system operation is estimated at 6.1 TWh which is roughly equivalent to 19 times the average electricity demand or average hourly load (av.h.l.$_{el}$). The discharge power capacity of EV batteries represents 1.3 TW, that is, 4 av.h.l.$_{el}$. The BEV state of charge is forced to be higher than 75\% at 5 a.m. every day (through $e_{n,s,t}$ in equation \ref{eq_storage}) to ensure that the batteries are full in the morning peak usage. This also restricts BEV demand to be shifted within a day and prevent EV batteries from becoming synoptic-scale or seasonal storage. The percentage of BEV connected to the grid at any time is inversely proportional to the transport demand profile, which translates into an average/minimum availability of 80\%/62\%. This approach is conservative compared to most of the literature. For instance, in \cite{circular_economy} the average parking time of the European fleet of vehicles is estimated at 92\%. The cost of the EV batteries is not included in the model since it is assumed that EV owners buy them to satisfy their mobility needs. This strong assumption is made to investigate the impact that a large exogenous storage capacity offered to the system at very low cost would have on its planning and operation.  Fuel Cell Electric Vehicles (FCEV) are not considered here since \cite{Brown_2018} showed that the extra flexibility provided by the use of hydrogen does not compensate the lower efficiency of fuel cells and they result in a more expensive system.

\paragraph{Individual Thermal Energy Storage (ITES)} \

Thermal energy can be stored in small water tanks with an estimated cost of 860 \EUR /m$^3$ \cite{Henning_2014}. Due to their reduced size, a non-negligible amount of the stored energy is lost per hour. The lost energy for individual tanks is estimated by $1-\exp(-\frac{1}{24\tau})$ with time constant $\tau=3$ days \cite{Lund_2016}. A temperature difference of 40K is assumed for thermal stores which corresponds to an energy density of 46.8 kWh$_{th}$/m$^3$ \cite{Brown_2018}.  \\

\paragraph {Central Long-term Energy Storage (CTES)} \

In district heating systems, large water tanks can be built close to the heat generators allowing seasonal thermal energy storage. For CTES a cost of 30 \EUR /m$^3$, a time constant of $\tau=180$ days and an energy density of 46.8 kWh$_{th}$/m$^3$ are considered \cite{Henning_2014}. These cost assumptions are in agreement with those reported in \cite{Lund_2016}. \\

\paragraph{Other storage technologies} \

There are some alternative storage technologies that are not included in this paper, among them, flywheels, supercapacitors, lead-acid batteries, Sodium-Sulfur, Vanadium redox-flow batteries, Compressed Air Energy Storage (CAES),  Concentrated Solar Power (CSP) comprising molten salts storage, heat storage in hot rocks or other underground heat storage. The mentioned technologies are at very different maturity level \cite{Schmidt_2017} and the cost predictions are particularly uncertain for those at an early stage.  In \cite{Cebulla_2017}, Vanadium redox-flow batteries are considered but the optimisation does not select them due to their higher cost compared to hydrogen and CAES also available in the model. In summary, to investigate the fundamental dynamics of storage in a highly renewable energy system we have restricted our model to a few technologies but still including a short-term and a long-term option for storing electricity and thermal energy. The possibility of producing synthetic methane and storing it is not included in the analysis.

\section{Methods} \label{sec_methods}
\subsection{Model} \label{sec_model}

\begin{figure}[ht!]
\centering
\includegraphics[trim=0 0cm 0 0,width=\linewidth,clip=true]{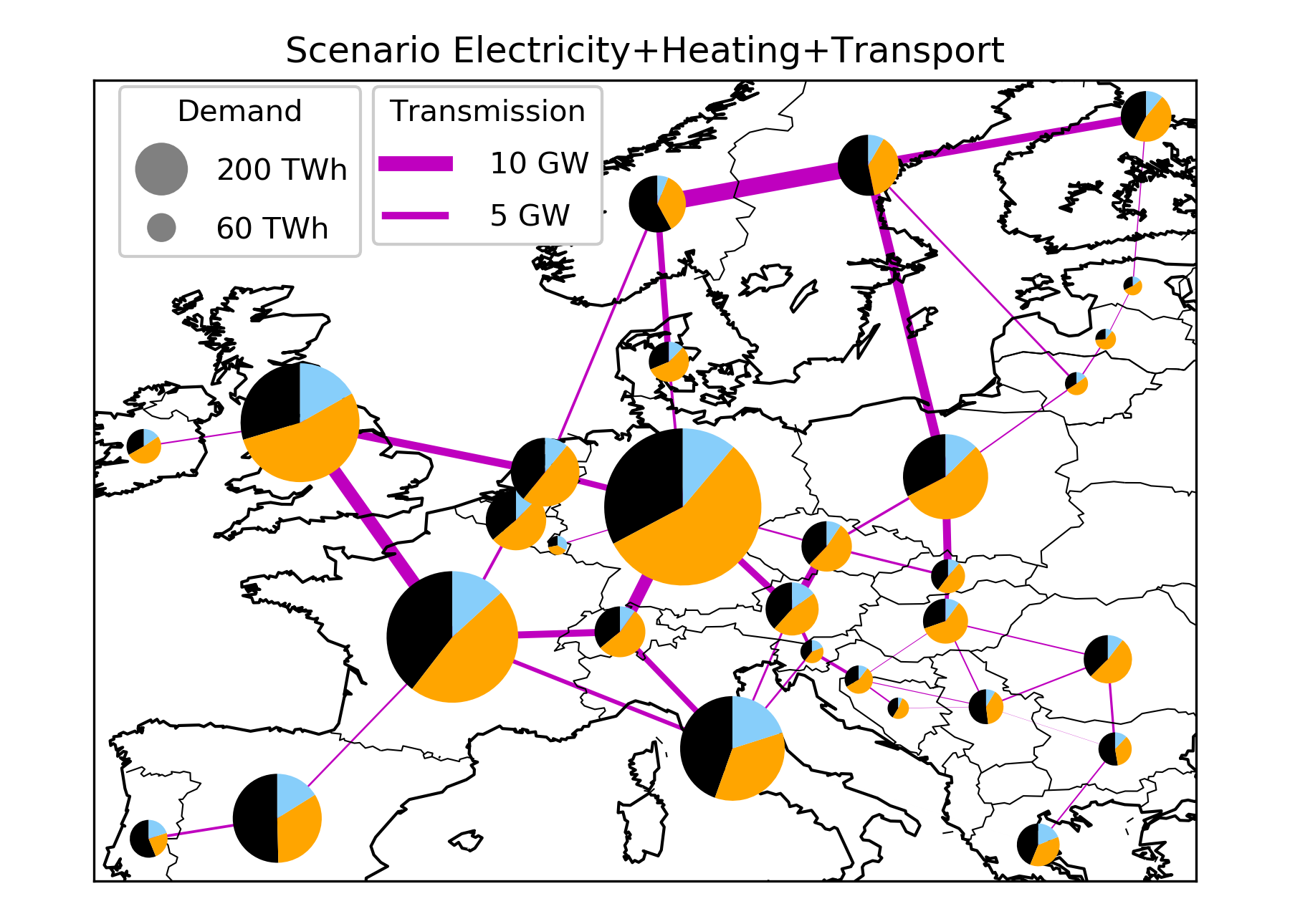}\\
\includegraphics[trim=0 0cm 0 0,width=\linewidth,clip=true]{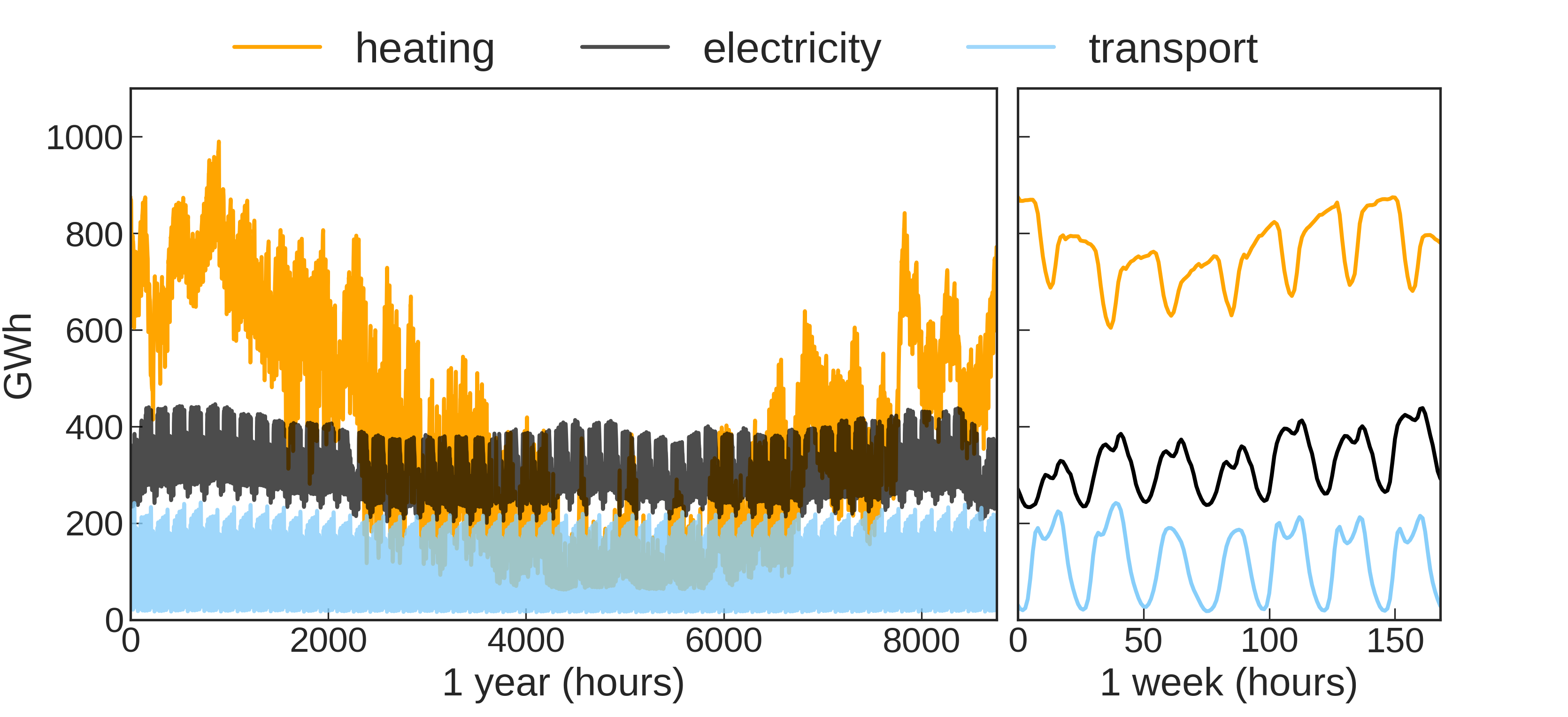}\\
\caption{(top) Spatial plot showing the energy demand per sector and country. (bottom) Europe-aggregated demand for the different sectors. Demands are estimated based on historical data, see \ref{anex_data}. } \label{fig_demands} 
\end{figure}

The role of storage technologies in the European energy system is investigated by using the PyPSA-Eur-Sec-30 model introduced in \cite{Brown_2018}. The model is built on the open-source framework Python for Power System Analysis (PyPSA) \cite{PyPSA} and uses hourly resolution for a full year. The network, shown in Figure \ref{fig_demands}, comprises 30 nodes, each of them represents a country of the 28 European Union member states as of 2018 excluding Malta and Cyprus but including Norway, Switzerland, Serbia, and Bosnia-Herzegovina. Neighbouring countries are connected through cross-border transmission lines. High Voltage Direct Current (HVDC) is assumed for the transmission lines, whose capacities can be expanded by the model if it is cost effective. Every country comprises three buses representing the electricity, heating, and transport sectors but not all the sectors are activated in every scenario under analysis (see Section \ref{sec_scenarios}). Within every country, buses are connected by energy converters, see Figure \ref{Fig_buses}.\\

A detailed description of the model including its mathematical formulation is provided in \ref{annex_pypsa_model}. For every scenario, the system configuration is optimised to minimise the total annualised system cost, Eq. (\ref{eq_objective}), subject to constraints, Eq. (\ref{eq_energy_balance}) to (\ref{eq_co2cap}), assuming perfect competition and foresight.  The model assumes long-term market equilibrium, that is, it ensures that the costs incurred by any optimised technology are exactly compensated by the market revenues. The inelastic loads in every sector, the hydroelectric capacities and, when the transport sector is activated, the storage capacity provided by EV batteries, are considered exogenous to the model and not optimised. By contrast, VRES generator capacities (onshore wind, offshore wind, and solar PV), conventional generator capacities (open cycle gas turbines, combined heat and power, gas boilers), converter capacities (heat pumps and resistive heaters), storage power and energy capacities (batteries and hydrogen for electricity and individual and central hot water tanks for heating), and transmission capacities are all optimised. The hourly operational dispatch of generators, converters, and storage units is also optimised.\\

The optimal system configuration is investigated for different CO$_2$ emissions caps, imposed by constraint (\ref{eq_co2cap}), and assuming independent or coupled sectors. In all the scenarios analysed, the current transmission capacity in Europe can be at the most doubled, \textit{i.e.}, $CAP_{LV}$ in Eq. (\ref{eq_cap}) is fixed at 62 TWkm. Only in Section \ref{sec_sensitivity} this constraint is released to analyse its effect on the optimal storage capacities. The average VRES generation (sum of wind and solar electricity) is assumed to be proportional to the average electricity load in every country, Eq. (\ref{eq_gamma}). In essence, we are demanding the different countries to be self-sufficient by generating in average what they consume. This constraint was not included in the previous analysis performed by Brown and coauthors \cite{Brown_2018, Brown_2019}. \\

A thorough description of the data used and the model assumptions for the different technologies is provided in \cite{Brown_2018}. In addition, we use here data collected for the year 2015 as described in \cite{Zhu_2019}. For the sake of conciseness, we refer the reader to \ref{annex_data} that includes a brief description of the data in every sector and Table \ref{tab_costs} that gathers the costs, lifetime, and efficiency values assumed for the different technologies.

\begin{table*}[!t]
\centering
	\begin{threeparttable}
		\caption{Costs, lifetime, and efficiency values assumed in the model. } \label{tab_costs}
		\centering
		\begin{footnotesize}
		\begin{tabularx}{0.65\textwidth}{lcccccc}
			\toprule
			Technology                 &Overnight   &Unit &FOM\tnote{b} &Lifetime & Efficiency & Source		\\
			&Cost\tnote{a}[\euro] &     &[\%/a] 		&[years]   	  &   & \\
			\midrule
			Onshore wind             &910         &kW\el  &3.3 &30  &	    & \cite{DEA_2016}\\
			Offshore wind             &2506        &kW\el  &3   &25  &		& \cite{DEA_2016} \\
			Solar PV utility-scale\tnote{c}  		   &425         &kW\el  &3 &25  &	 	&  \cite{Vartiainen_2017} \\
			Solar PV rooftop\tnote{c}  		   &725         &kW\el  &2 &25  &	 	&  \cite{Vartiainen_2017} \\			
			Hydro reservoir\tnote{f} & 2000 & kW\el & 1 & 80 & & \cite{Schroeder_2013} \\			
			Run-of-river\tnote{f} & 3000 & kW\el & 2 & 80 & & \cite{Schroeder_2013} \\							
			OCGT\tnote{d}              &560         &kW\el  &3.3 &25  &0.39 &  \cite{DEA_2016,Schroeder_2013}       				                   \\
			CHP\tnote{d}       	   &600         &kW\th  &3.0 &25  &0.47    &    	\cite{Henning_2014}			                     \\
			Gas boiler\tnote{d,e}      &175/63      &kW\th  &1.5 &20  &0.9    &     	\cite{Palzer_thesis}			    \\
			Resistive heater           &100         &kW\th  &2   &20  &0.9   		& \cite{Schaber_2013}      \\
			Heat pump\tnote{e}         &1400/933    &kW\th  &3.5 &20  & $\approx$ 3-4\tnote{e} 	  &	 \cite{Henning_2014, Palzer_thesis}	\\
			HVDC lines	  			   &400			&MWkm	&2	 &40  &		   			& \cite{Hagspiel_2014}						\\
			HVDC converter pair	  			   &150			&kW	&2	 &40  &		   			& \cite{Hagspiel_2014}						\\
			\bottomrule
		\end{tabularx}
		\end{footnotesize}
		\begin{tablenotes}
			\footnotesize
			\item [a] The overnight costs are annualised assuming a discount rate of 0.07.
			\item [b] Fixed Operation and Maintenance (FOM) costs are given as a percentage of the overnight cost per year.
			\item [c] 50\% of the installed capacities are rooftop-mounted systems and 50\% utility-scale power plants, 4\% and 7\% discount rates have been assumed respectively. 
			\item [d] CHP fuel is gas. The fuel cost of OCGT, CHP, and gas boiler is 21.6 \EUR/MWh$_{th}$. Efficiency for CHP reported here is calculated assuming they work in condensing mode but a complete model for CHP operation is included, see \cite{Brown_2018}.
			\item [e] Gas boilers and heat pumps have different costs for individual (numbers in front) and centralised (numbers behind) systems. The efficiency of heat pumps, also known as the Coefficient of Performance (COP), is calculated for every hour using the ambient temperature, see \cite{Brown_2018}.
			\item [f] \footnotesize{Reservoir, run-of-river and PHS are exogenous to the system. The capacities in every country are fixed and they are considered to be fully amortised. }
		\end{tablenotes}
	\end{threeparttable}
	\centering
\end{table*}

\subsection{Scenarios under analysis} \label{sec_scenarios}

We use the network model PyPSA-Eur-Sec-30 with four different scenarios:

\begin{itemize}
\item	Electricity
\item	Electricity + Heating 
\item	Electricity + Transport 
\item	Electricity + Heating + Transport
\end{itemize}

Scenario Electricity includes only current electricity demand (see \ref{annex_data}). When a new sector is included, the associated demand, as well as the generation and storage technologies, are added to the model. The storage technologies included in every scenario are shown in Table \ref{tab_stores} and their characteristics are summarised in Table \ref{tab_stores_characteristics}. \\

\begin{table*}
	\centering
		\caption{Available storage technologies in every scenario. Characteristics of the different storage technologies are shown in Table \ref{tab_stores_characteristics}. } \label{tab_stores}
		\
		\begin{footnotesize}
    \begin{tabular}{p{7cm}p{2cm}p{2cm}p{2cm}p{2cm}}
		\hline
     & Electricity & Electricity+ Heating & Electricity+ Transport & Electricity+ Transport+ Electricity \\
		\hline
		Batteries & \OK & \OK & \OK & \OK \\
    Hydrogen    & \OK & \OK & \OK & \OK\\
    Pumped hydro storage (PHS) & \OK & \OK & \OK & \OK  \\
    EV batteries &  &  & \OK & \OK \\
    Individual short-term thermal energy storage (ITES) &       &   \OK    &      & \OK \\
    Central long-term thermal energy storage (CTES) &       &     \OK  &      & \OK \\
		\hline
   
		\end{tabular}
		\end{footnotesize}
\end{table*}

\section{Results and discussions} \label{sec_results}

\subsection{Cost-effective energy and power capacities for storage technologies} \label{sec_capacities_evolution}

\begin{figure*}[ht!]
\centering
\includegraphics[width=0.9\textwidth]{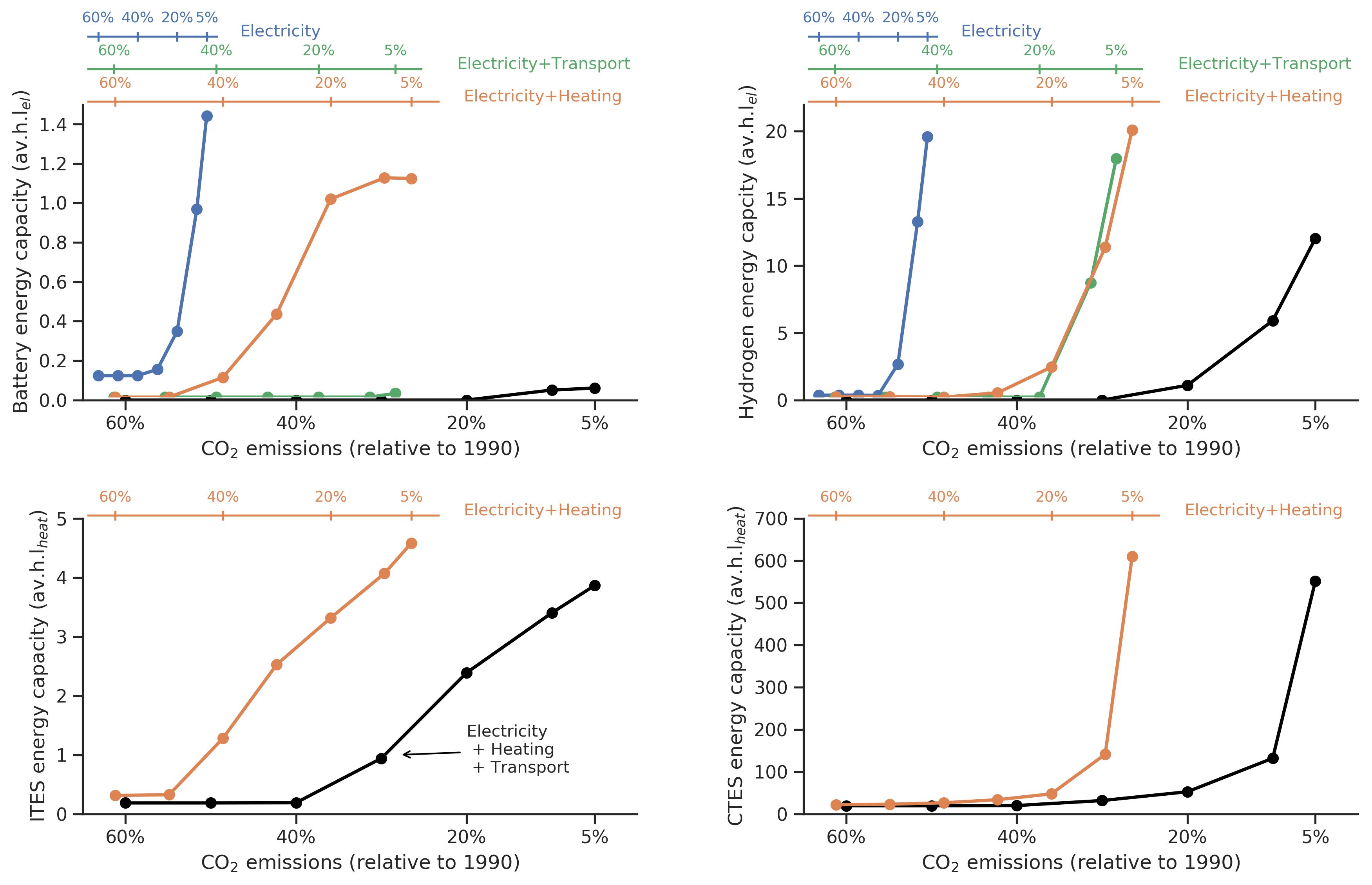}\\
\caption{Cost-effective energy capacity as a function of CO$_2$ emissions for different storage technologies: electric batteries (top left), hydrogen storage (top right), individual short-term thermal energy storage, ITES (bottom left), and central long-term thermal energy storage, CTES (bottom right). Values are shown normalised by the average hourly load in the Electricity sector av.h.l.$_{el}$ and the average hourly heating demand av.h.l.$_{heat}$. The top horizontal axes indicate the CO$_2$ emissions relative to Electricity (blue), Electricity+Transport (green), and Electricity+Heating (orange) levels in 1990. The bottom horizontal axis indicates the CO$_2$ emissions relative to the three-sectors-aggregated emissions in 1990. Two additional storage technologies whose energy and power capacities are fixed, not subject to optimisation, are present in some of the scenarios: pumped hydro storage and EV batteries (see Table \ref{tab_stores}). } \label{fig_capacities} 
\end{figure*}

\begin{figure*}[ht!]
\centering
\includegraphics[width=0.8\textwidth]{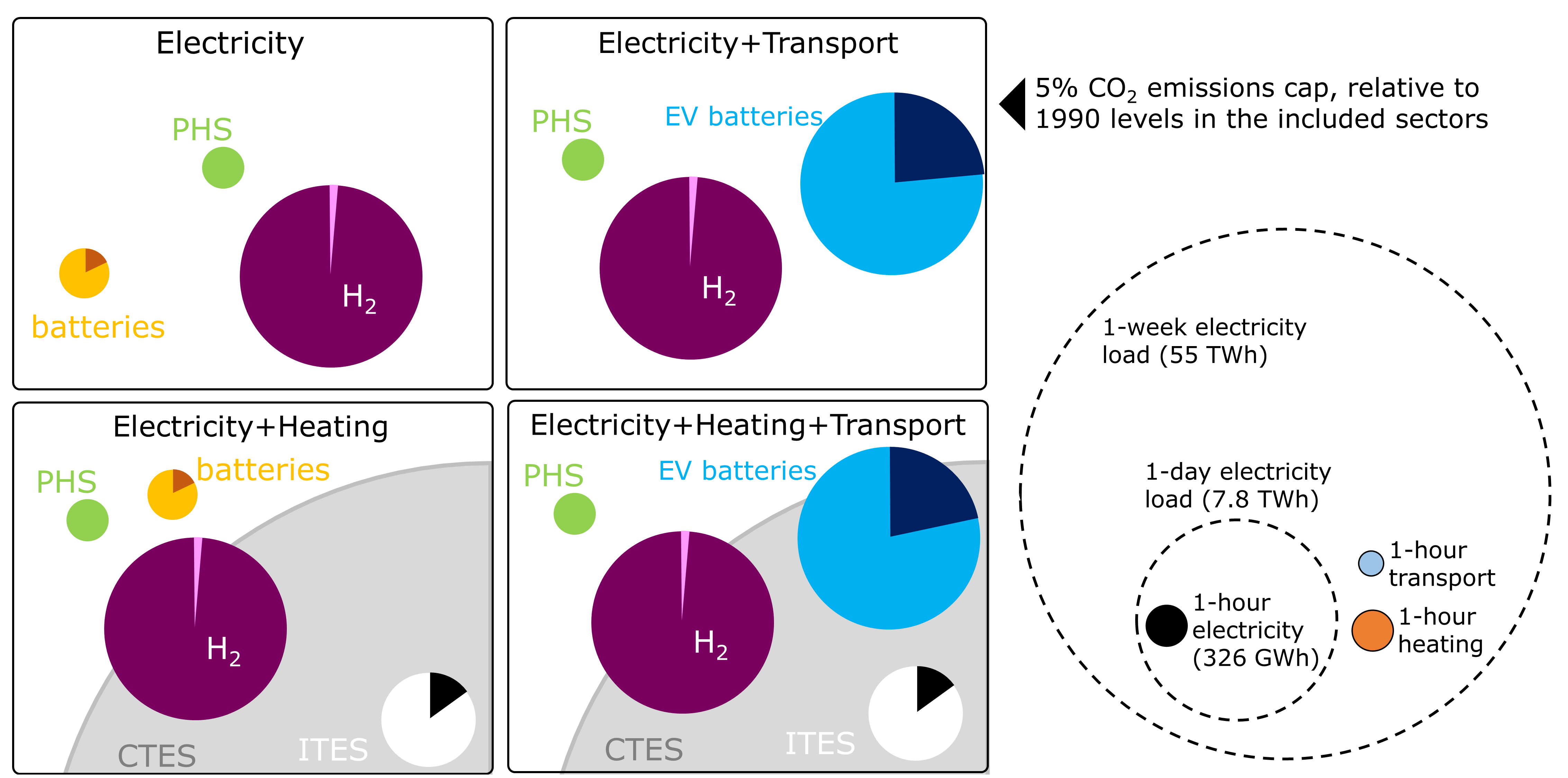}\\
\caption{The size of the circles represents the cost-effective energy capacity for electric batteries, hydrogen storage, individual (ITES) and central long-term thermal energy storage (CTES). For comparison, the average hourly, daily, and weekly electricity demand, the average hourly heating and transport demand, and the energy capacity of PHS and EV batteries are shown. For every storage technology circle, the marked sector represents the energy that can be discharged in one hour. In all the scenarios, CO$_2$ emissions cap is set to 5\% of 1990 levels in the included sectors.} \label{fig_bubbles} 
\end{figure*}

\begin{figure}[!h]
\centering
	\includegraphics[trim=0 0cm 0 0,width=0.8\linewidth,clip=true]{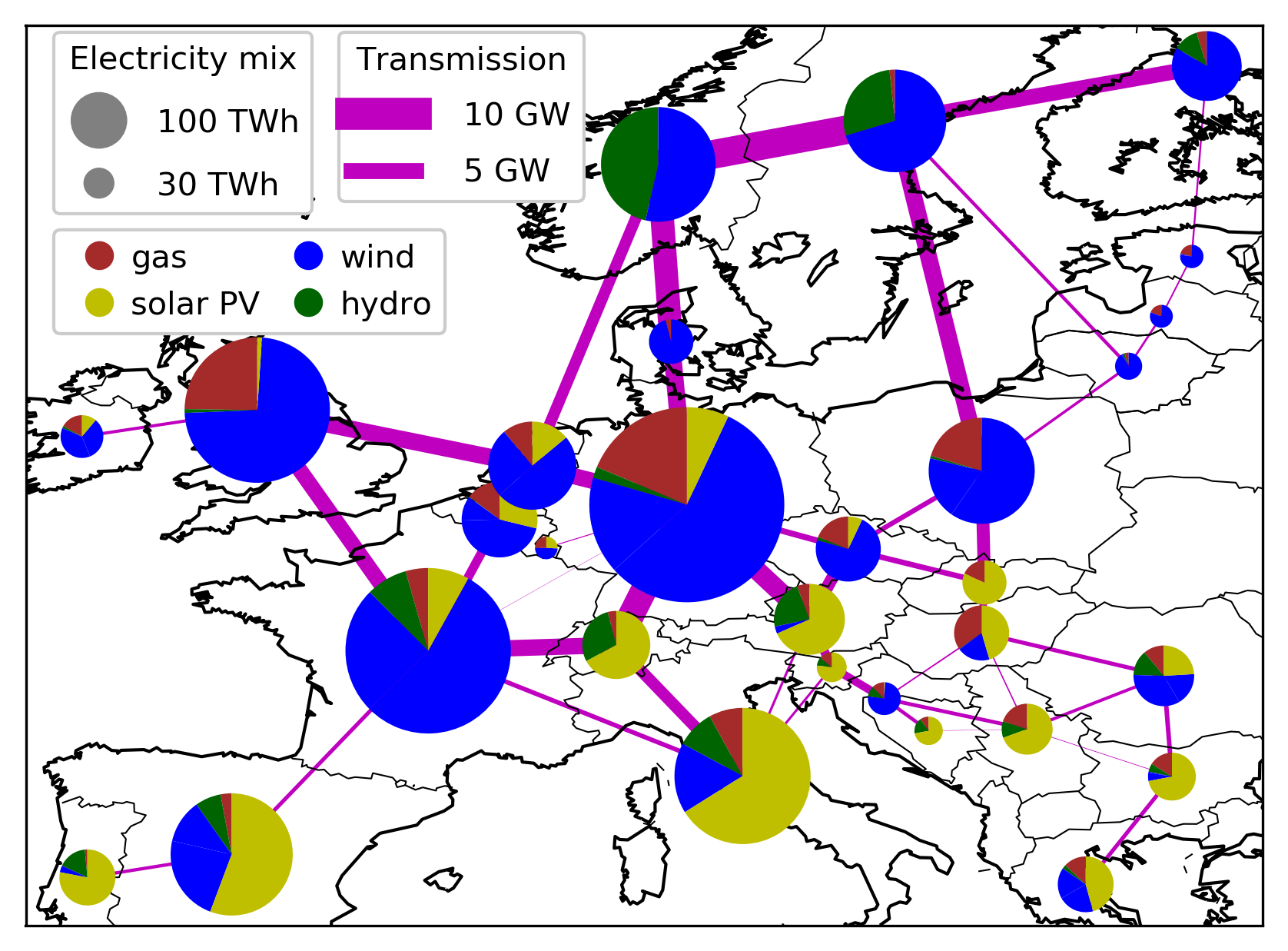}
	\includegraphics[trim=0 0cm 0 0,width=0.86\linewidth,clip=true]{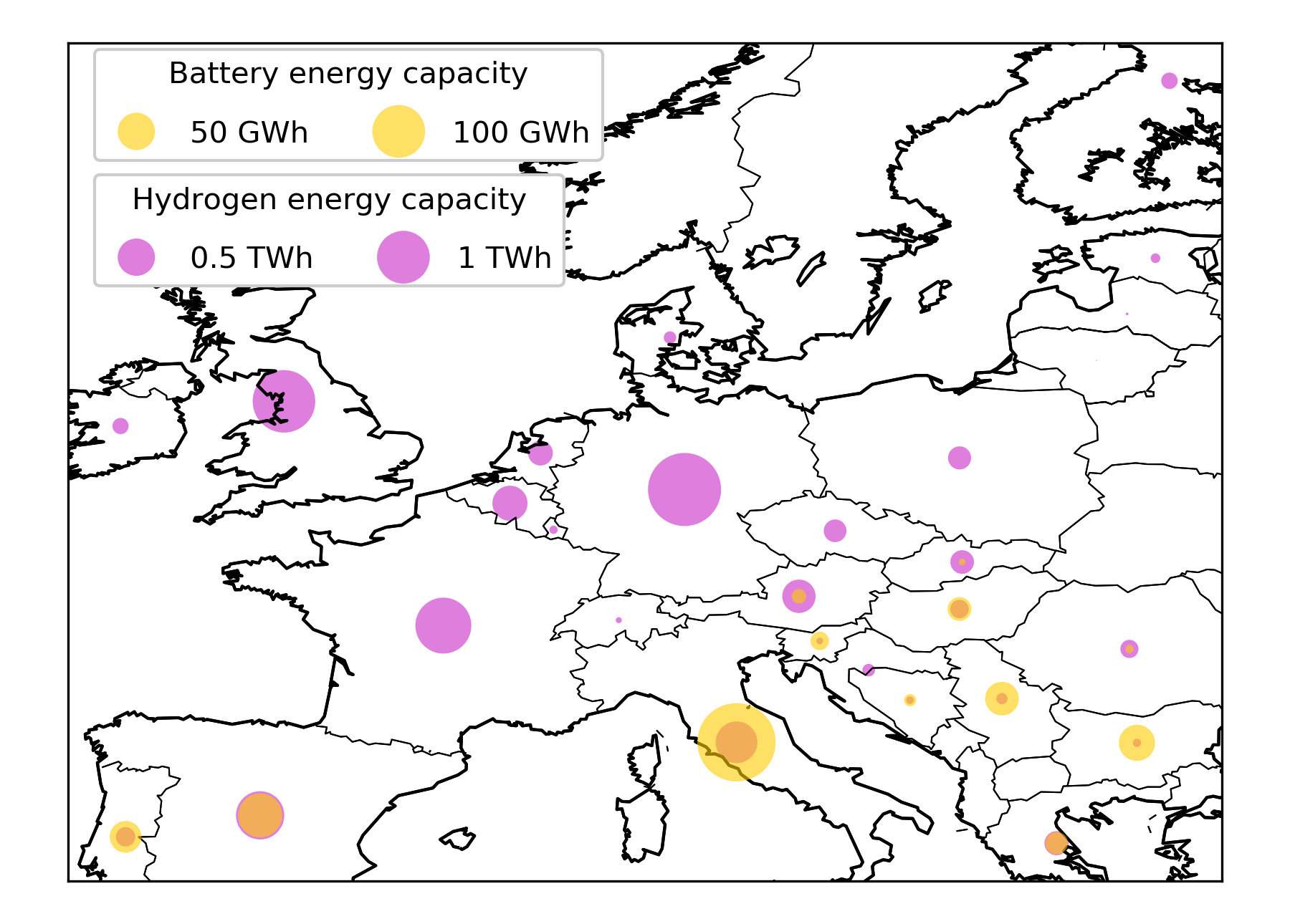}
	\includegraphics[trim=0 0cm 0 0,width=0.86\linewidth,clip=true]{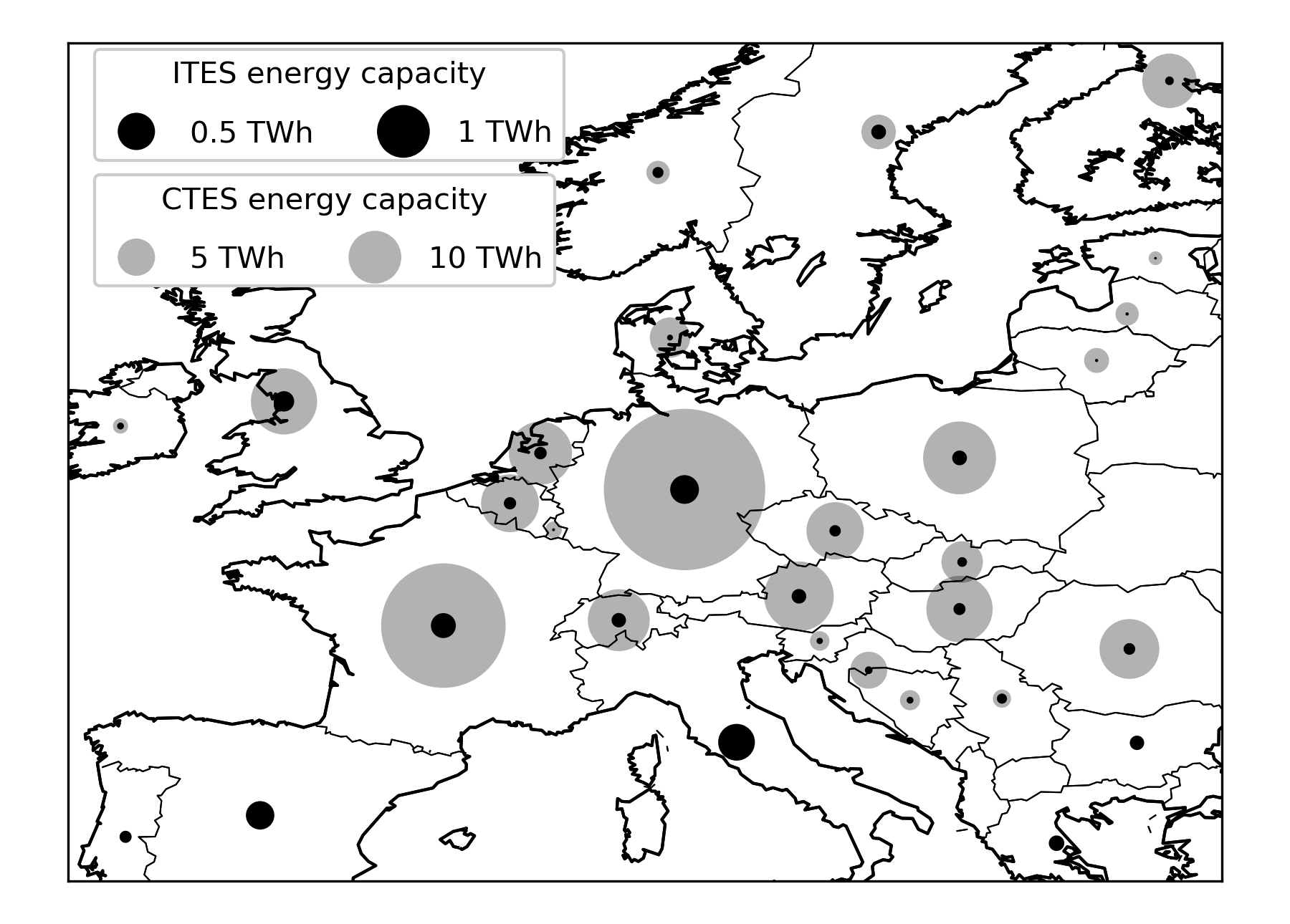}	
\caption{Spatial plots showing the shares of electricity generation (top) in the scenario Electricity + Heating. Energy capacities for batteries and hydrogen (middle), individual (ITES) and central (CTES) thermal energy storage (bottom). CO$_2$ emissions cap equal to 5\% of 1990 level is imposed.} \label{fig_spatial plots} 
\end{figure}

\begin{figure}[ht!]
\centering
\includegraphics[width=0.9\columnwidth]{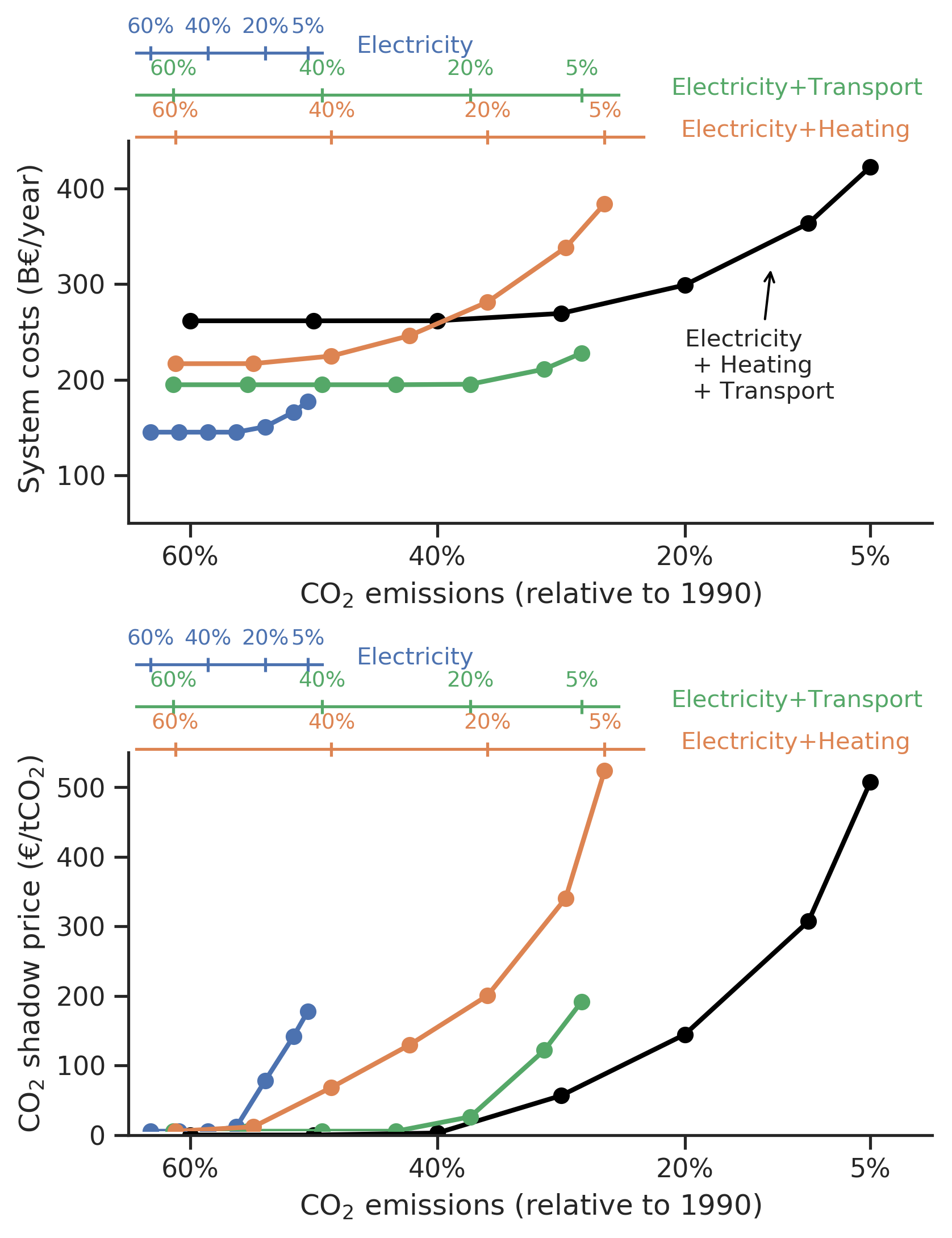}\\
\caption{(top) Annualised system cost and (bottom) CO$_2$ shadow price as a function of the CO$_2$ emissions cap relative to 1990. The top horizontal axes indicate the CO$_2$ emissions relative to Electricity (blue), Electricity+Transport (green), and Electricity+Heating (orange) levels in 1990. The bottom horizontal axis indicates the CO$_2$ emissions relative to the three-sectors-aggregated emissions in 1990.} \label{fig_system_cost}
\end{figure}

Figure \ref{fig_capacities} plots the Europe-aggregated storage energy capacities included in the cost-optimal configuration of the system under CO$_2$ emissions caps ranging from 60\% to 5\%, relative to 1990 levels. As a graphical summary, Figure \ref{fig_bubbles} provides an overview of the required aggregated storage energy capacities for 5\% emissions cap. The energy capacities are normalised by the Europe-average electricity load (av.h.l.$_{el}$) which in 2015 accounted for 326 GWh. Available PHS is fixed and accounts  for an aggregated energy capacity of 285 GWh (0.87 av.h.l$_{el}$) and power capacity of 47.5 GW.\\

We start by discussing the scenario Electricity (blue lines in Figure \ref{fig_capacities}). The cost-effective energy capacities for batteries and hydrogen storage remain almost constant until the CO$_2$ cap reaches 30\% of 1990 emissions in that sector (upper blue horizontal axis) and then increase steeply. The reason behind the constant values for CO$_2$ caps higher than 30\% is the fact that the CO$_2$ emissions constraint is not binding so the optimal system configuration is the same, this is further discussed below. In 1990, electricity generation was responsible for 1510 MtCO$_2$, land-based transport for 784 MtCO$_2$, and heating in the residential and service sectors for 723 MtCO$_2$ \cite{ODYSSEE}. Consequently, reducing CO$_2$ emissions in the Electricity sector to 5\% of 1990 values  and assuming that heating and transport sectors remain unchanged, implies that the emissions of the three sectors are 53\% of the aggregated 1990 values. This is shown in the lower horizontal axis of Figure \ref{fig_capacities}. \\

The optimal batteries energy capacity for the Electricity scenario and 5\% CO$_2$ emissions cap is equivalent to 1.4 av.h.l.$_{el}$ and the batteries power capacity, which is independently optimised in the model, accounts for 0.25 av.h.l.$_{el}$ This means that batteries discharges in 5.6 hours at full power. This result agrees with the 6 hours discharge time identified by Rasmussen and coauthors \cite{Rasmussen_2012} using the weather-driven modelling approach. The reason is that the expensive electric batteries, with high round-trip efficiency, are mainly used to counterbalance solar generation and, consequently, their optimal discharge time is strongly influenced by the daily pattern of solar generation.\\

The cost-effective hydrogen storage energy and power capacities are equivalent to 19.4 and 0.48 av.h.l.$_{el}$ respectively. At maximum power, hydrogen stores would be depleted roughly in 2 days. This agrees with the preliminary analysis performed in Section \ref{sec_storage_technologies}: batteries are used for short-term storage while hydrogen is preferred for the long term. The dispatch patterns will be further investigated in Section \ref{dispatch_pattern}. It is worth mentioning that the cost assumed for the hydrogen stores (Table \ref{tab_stores_characteristics}) corresponds to overground steel tanks, so no geographical constraint is included to limit its deployment. However, hydrogen can also be stored in salt caverns at a lower cost \cite{Steward_2009}. The storage capacity of the existing salt caverns in Europe is significantly larger than the required energy capacity estimated here \cite{Staffell_2019, Gilhaus_2007, Cebulla_2017}. \\

If we look now at the Electricity+Transport scenario (green lines in Figure \ref{fig_capacities}), it can be observed that it does not include electric batteries in the optimum configuration. The large energy capacity provided by EV batteries makes them unnecessary (compare the yellow circle representing the cost-effective batteries energy capacity in the scenario Electricity, Figure \ref{fig_bubbles}, and the blue circle representing the energy capacity provided by the EV batteries). It must be pointed out here that the cost of EVs is not included in the optimisation, but the corresponding EV batteries are just assumed to be available in the scenarios including the transport sector. Moreover, in as much as the energy and power capacity storage provided by EV batteries is significantly larger than the optimised values for electric batteries in the scenario Electricity, this result is not very sensitive to the assumed hypothesis, \textit{i.e.}, half of the European cars provide vehicle-to-grid services. The maximum required hydrogen storage is similar to that obtained in the scenario Electricity and, what is more relevant, the three sectors aggregated CO$_2$ emissions curb to 45\% before a large energy capacity of hydrogen storage is needed. In essence, the increase in demand caused by electrifying transport delays the moment where things start to become difficult.\\

The scenario Electricity+Heating installs a large energy capacity for central long-term thermal energy storage (CTES), that is, large water tanks connected to district heating systems. Large CTES capacity appears at CO$_2$ emissions equal to 30\% (upper orange horizontal axis in Figure \ref{fig_capacities}) and increases significantly reaching an energy capacity roughly equivalent to 23 days of average heat demand, av.h.l$_{heat}$. This represents a total of 4.8 billion cubic metres of hot water, which is equivalent to 20 cubic metres per citizen. Individual thermal energy storage appears for CO$_2$ emissions equal to 50\% and reaches only 5 av.h.l$_{heat}$ since the low time constant assumed for the thermal decay of ITES, $\tau$=3 days, prevent them from becoming seasonal storage. ITES accounts for 0.14 cubic metres per citizen. The cost-effective batteries and hydrogen storage energy capacity is similar to the Electricity scenario but its emergence is delayed. Figure \ref{fig_spatial plots} depicts the spatial distribution of cost-effective storage energy capacities per country in the scenario Electricity+Heating. As expected, larger batteries capacities are obtained in countries with high solar penetration and larger hydrogen storage energy capacities correlate with wind generation. This is in agreement with the layout obtained for the power system in \cite{Schlachtberger_2017}. The heating sector included here installs large CTES energy capacities in those countries where district heating is allowed. ITES is present in all the countries, associated with individual heating systems in rural areas, and it is particularly important in southern countries where individual heating systems are also employed in urban areas.\\

Finally, the scenario Electricity+Heating+Transport (black lines in Figure \ref{fig_capacities}) shows the combined effects of both EV batteries and thermal energy storage. One remarkable result is that, by coupling the three sectors, large storage energy capacities are not necessary until the global CO$_2$ emissions cap reaches 20\%. \\

Figure \ref{fig_system_cost} shows the evolution of the annualised system cost as a function of the CO$_2$ limit for the different scenarios. Under our costs assumptions, the CO$_2$ constraint, Eq. (\ref{eq_co2cap}), is binding only from a certain CO$_2$ emissions cap that depends on the sectors included in the scenario. For higher CO$_2$ emissions, the system optimal configuration remains the same. This can be seen in Figure \ref{fig_system_cost} where the Lagrange/Karun-Kush-Tucker multiplier, $\mu_{CO_{2}}$ of constraint (\ref{eq_co2cap}), is plotted. $\mu_{CO_{2}}$  represents the shadow price of CO$_2$, \textit{i.e.}, the additional price that should be added for every unit of CO$_2$ to achieve the CO$_2$ reduction target in an open market. The carbon price required when the heating sector is included is significantly higher than when the electricity sector is independently decarbonised. Roughly speaking, carbon price represents the economic penalty that should be added to an emitting technology in order to alter the existing merit order so that the optimiser selects the next technology with lower emissions. While in the electricity sector low carbon prices are enough to prioritise renewable generation over OCGT, this is not the case for the heating sector in which high CO$_2$ prices are needed to make heat pumps more attractive when compared to gas boilers. This is why scenarios including the heating sector show a similar CO$_2$ shadow price. The CO$_2$ price estimated here is heavily impacted by the fact that the model does not include biomass which would represent an alternative option to gas boilers requiring a lower CO$_2$ price to become cost competitive. Consequently, the CO$_2$ prices shown here should not be understood as a policy recommendation but as an indication of the enormous challenge that decarbonising the heating sector represents with the current costs structure. The reader is referred to \cite{Zhu_2019} for a thorough analysis of the impact of CO$_2$ prices in a coupled electricity-heating system in Europe.

\subsection{Storage dispatch patterns} \label{dispatch_pattern}

\begin{figure}[ht!]
\centering
\includegraphics[width=\columnwidth]{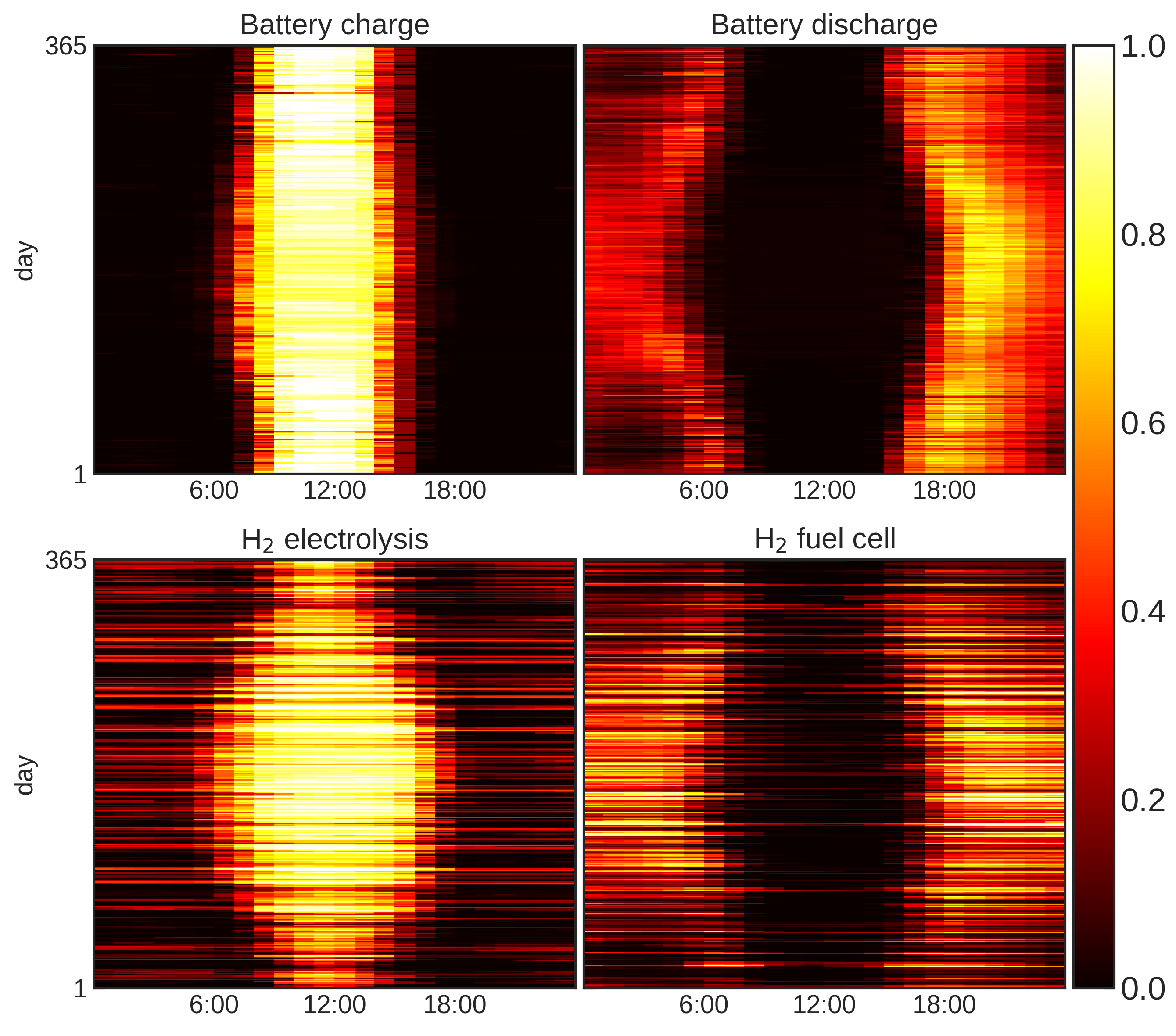}\\
\caption{Europe-aggregated charging and discharging power for batteries and hydrogen storage for the scenario Electricity+Heating and CO$_2$ emissions limited to 5\% of 1990. For every hour the charging/discharging power is normalised by the Europe-aggregated power capacity. } \label{fig_heatmaps}
\end{figure}

\begin{figure}[ht!]
\centering
\includegraphics[width=0.8\columnwidth]{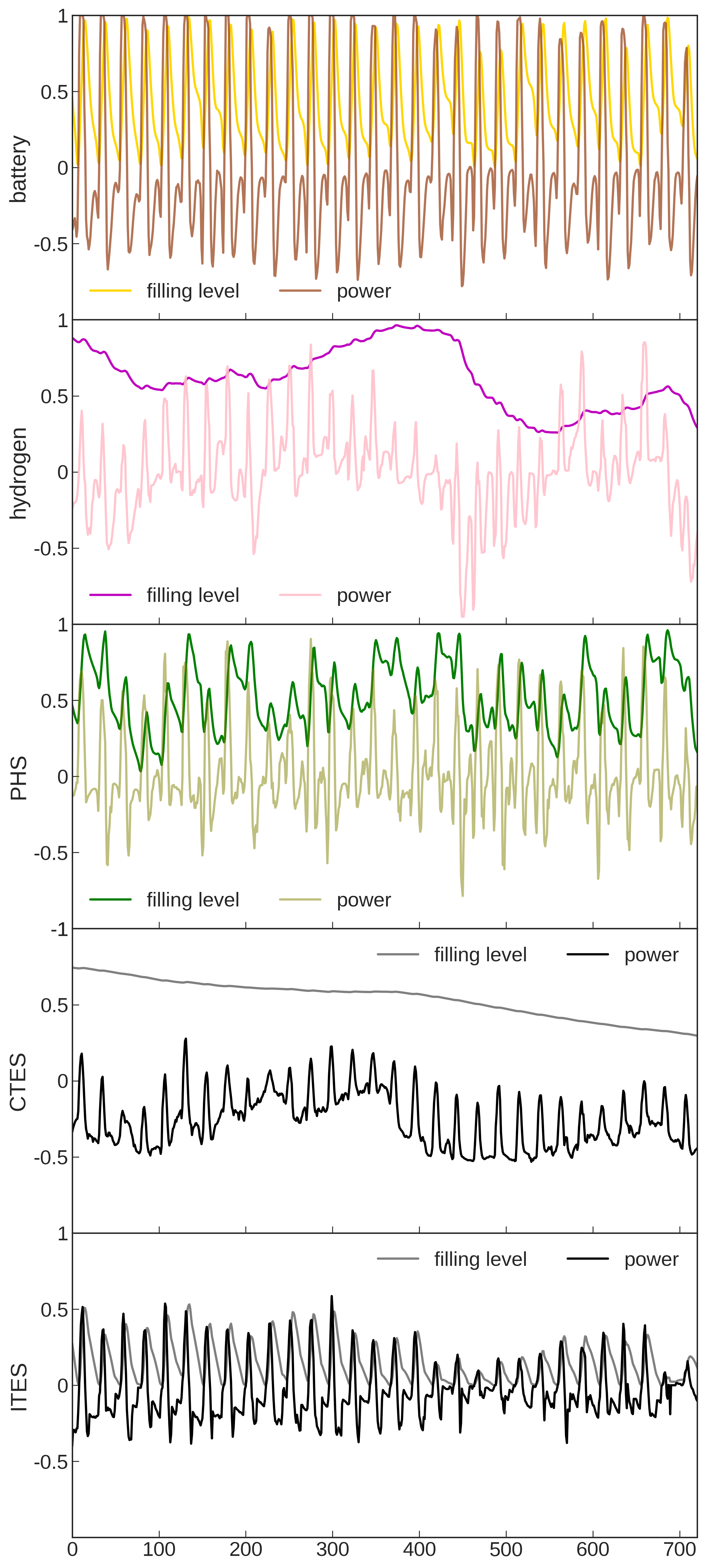}\\
\caption{Europe-aggregated normalised charging and discharging power and filling level of the storage technologies in the scenario Electricity+Heating for 5\% CO$_2$ emissions. Hourly values throughout January are shown.} \label{fig_level_power}
\end{figure}

\begin{figure*}[ht!]
\centering
\includegraphics[width=0.7\textwidth]{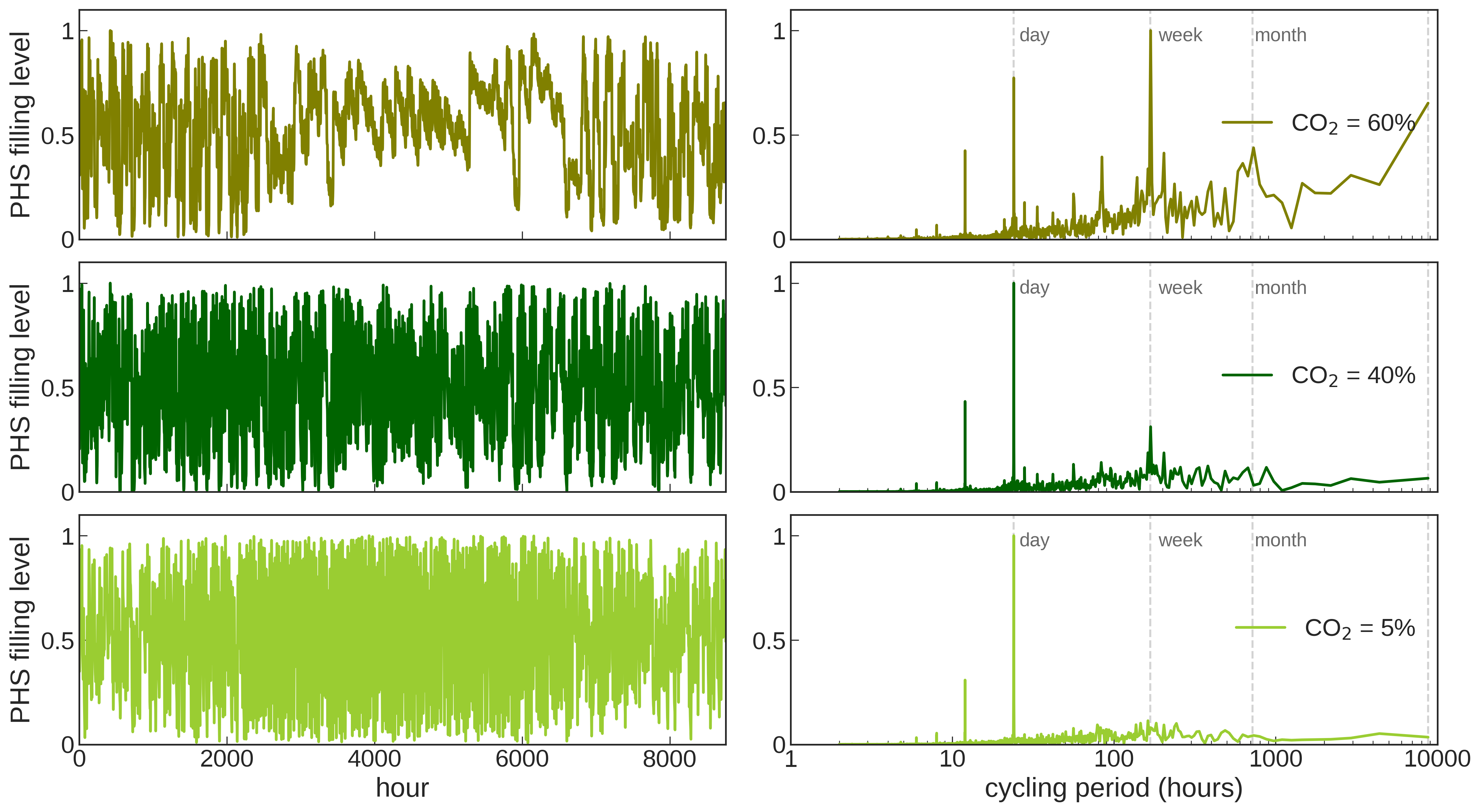}\\
\caption{Europe-aggregated hourly filling level of pumped hydro storage (PHS) for the scenario Electricity+Heating and CO$_2$ emissions targets equal to 60\%, 40\% and 5\%, relative to 1990 levels. The plots on the right show the Fourier power spectrum of the time series in the left plots. Vertical grey dashed lines indicate cycling periods corresponding to one day, week, month, and year.} \label{fig_fourier_PHS}
\end{figure*}

\begin{figure*}[ht!]
\centering
\includegraphics[width=\textwidth]{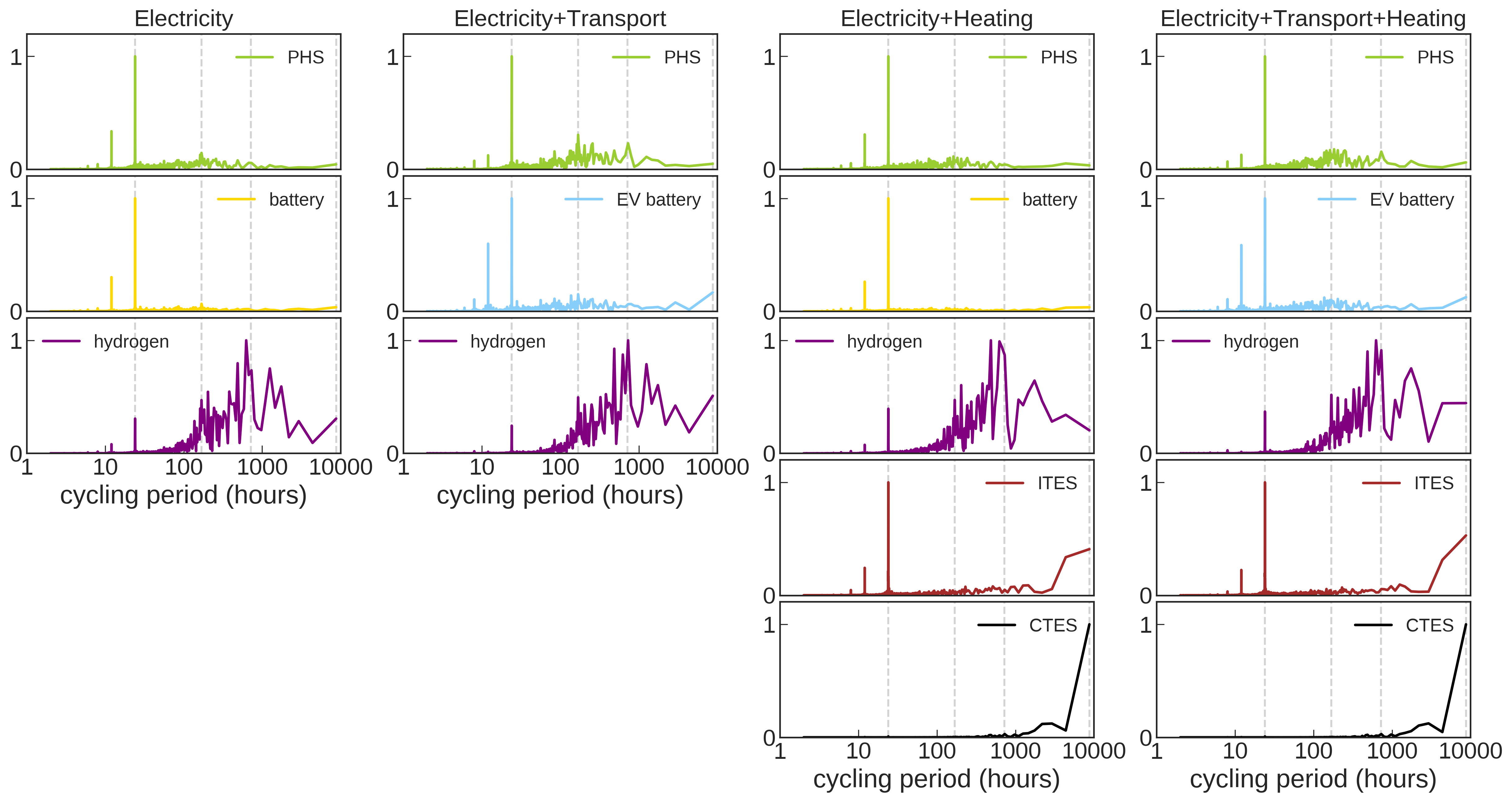}\\
\caption{Fourier power spectra of the Europe-aggregated hourly filling level of the storage technologies in the scenarios under analysis for CO$_2$ emissions limited to 5\% relative to 1990 levels. To compute the Europe-aggregated filling level for every hour, the sum of the stored energy in Europe is divided by the sum of the storage energy capacities. Vertical grey dashed lines indicate cycling periods corresponding to one day, week, month, and year.} \label{fig_fourier_sector_coupling}
\end{figure*}
 
The charge and discharge patterns of storage technologies mainly depend on the time series of the mismatch energy, \textit{i.e.}, the difference between VRES generation and demand in every hour. The results included in this section show how these patterns are strongly influenced by the renewable penetration, which is directly linked to the CO$_2$ emissions cap in every scenario, and the presence of alternative storage technologies.  Let us start describing the general trends in the storage dispatch patterns. Figure \ref{fig_heatmaps} depicts the heatmaps for the Europe-aggregated normalised charging and discharging power of batteries and hydrogen storage at every hour in the scenario Electricity+Heating under 5\% CO$_2$ emissions cap. The Europe-aggregated normalised power is calculated, for every storage technology, by dividing the sum of the charge/discharge power in all the countries by the sum of their maximum power capacities. Although, as a general trend, both storage technologies charge during the day and discharge throughout the night, batteries show a sharper diurnal pattern while hydrogen storage shows larger seasonal variability including several contiguous days in winter of permanent electricity production through fuel cells to compensate for low VRES generation, as well as few days of stable hydrogen production through electrolysis to store VRES surplus. \\

Figure \ref{fig_level_power} plots the Europe-aggregated storage filling level and charge/discharge normalised power throughout January. This month is chosen as an example to depict the hourly profiles, but a comprehensive comparison based on the Fourier power spectra and using the whole year is presented later. The Europe-aggregated filling level is calculated, for every storage technology,  by dividing the sum of the stored energy in all the countries by the sum of their maximum energy capacities. The short-term and long-term role adopted by batteries and hydrogen storage respectively is determined by their energy and power capacity costs as anticipated in Section \ref{sec_storage_technologies}. The batteries dispatch profile depicted in Figure \ref{fig_level_power} shows positive values that approach 1, while negative values do not get lower than -0.7. This is a consequence of the model assumption in which charge and discharge capacities for batteries are requested to be equal. It also indicates that the charging phase determines the cost-effective power capacity for batteries in order to be able to store solar surplus in the middle of the day effectively. Then, the stored energy is discharged throughout the night without using the maximum available power capacity. This behaviour was already identified in \cite{Cebulla_2017}. PHS also shows a strong diurnal pattern and makes use of higher capacities when charging the storage, again influenced by solar midday surplus. For hydrogen storage, the capacities for the production of hydrogen via electrolysis, the storage in steel tanks, and the generation of electricity with fuel cells are independently optimised. The evolution of the filling level for hydrogen storage shows a lower frequency. ITES charges and discharges daily while CTES shows a smooth constant discharge throughout January. \\

We further investigate now the time series of storage filling level and start by analysing the behaviour of PHS in the scenario Electricity+Heating. The case of PHS is particular. While the storage and power capacities for other storage technologies are optimised together with their hourly dispatch, PHS capacities are assumed to be fixed and equal to the cumulative installed values in 2013 (see Table \ref{tab_stores_characteristics}). Figure \ref{fig_fourier_PHS} depicts the time evolution of the Europe-aggregated filling level of PHS for different CO$_2$ emissions caps and, consequently renewable penetrations.  As the renewable penetration increases, PHS dispatches energy more frequently. The right plots in Figure \ref{fig_fourier_PHS} show the Fourier power spectrum of the filling level time series where the predominant dispatching frequencies can be appreciated. For 5\% CO$_2$ emissions, the PHS daily frequency increases compared to the dispatch under 60\% CO$_2$ emissions constraint, and weekly, monthly and seasonal components disappear. For hydrogen storage, not shown in the figure, decreasing CO$_2$ emission cap from 60\% to 5\%, reduces the seasonal component and increases the predominance of weakly and monthly frequencies. The change of behaviour of PHS and hydrogen storage resembles the analysis of backup flexibility classes performed by Schlachtberger \textit{et al.} \cite{Schlachtberger_2016}. These authors found that highly flexible backup is more necessary as renewable penetration increases. \\

To analyse how the available storage technologies compete to provide flexibility at different time scales, Figure \ref{fig_fourier_sector_coupling} depicts the Fourier power spectra of the Europe-aggregated filling level of the available storage technologies for the four scenarios under analysis. In all the plots a CO$_2$ emissions cap equal to 5\% of 1990 levels (aggregated for the sectors included in every scenario) is assumed. For the Electricity scenario, electric batteries cycles daily counterbalancing the generation pattern of solar PV, while hydrogen storage discharge every 1-2 weeks influenced by the wind generation synoptic time scale. PHS behaves as daily storage complementing the role of electric batteries. For the Electricity+Transport scenario, as already mentioned, the short-term storage provided by EV batteries makes stationary electric batteries unnecessary. It should be reminded here that the constraint imposing almost full capacity in the early mornings for EV batteries (Section \ref{sec_storage_technologies}) restrains this storage from becoming seasonal. The presence of EV batteries, together with the additional electricity demand from EVs, also affects PHS behaviour increasing the lower cycling frequencies. For hydrogen storage, dominant frequencies remain the same. For the Electricity+Heating scenario, due to the low capital cost of CTES, the optimal system configuration includes a large energy capacity that behaves as seasonal storage, charging in summer and discharging in winter. ITES shows both daily and seasonal frequencies. Finally, for the scenario Electricity+Transport+Heating, individual ITES, PHS, and EV batteries show daily cycling, hydrogen dominant frequencies remain in the week-month period, and CTES shows a strong seasonal frequency. \\

\subsection{Market revenues of PHS and hydroelectricity} \label{revenues_hydro}

\begin{figure}[ht!]
\centering
\includegraphics[width=0.8\columnwidth]{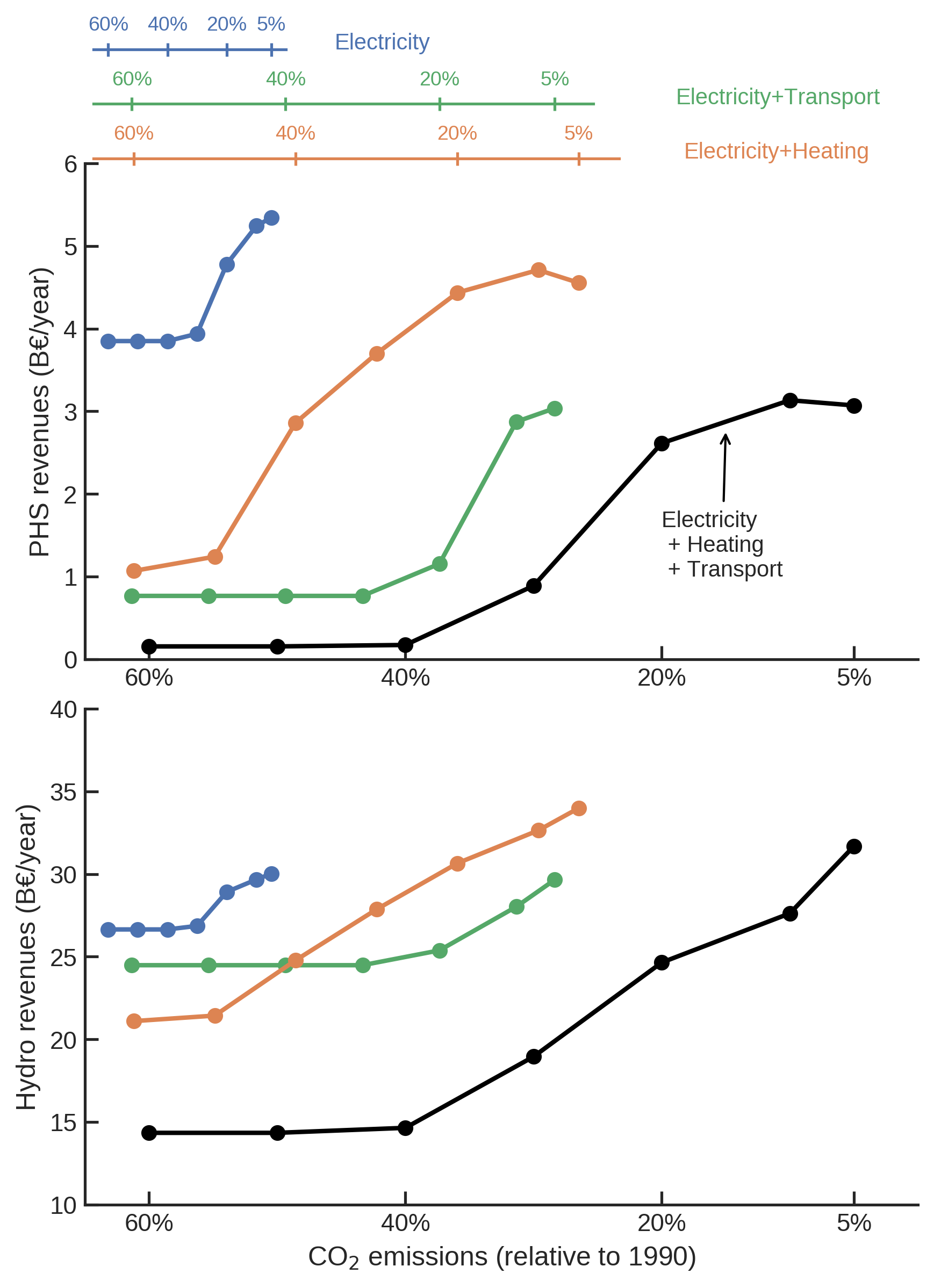}\\
\caption{Pumped hydro storage (top) and reservoir hydroelectric (bottom) annual market revenues vs CO$_2$ emissions, relative to the sectors included in every scenario in 1990. The top horizontal axes indicate the CO$_2$ emissions relative to Electricity (blue), Electricity+Transport (green), and Electricity+Heating (orange) levels in 1990. The bottom horizontal axis indicates the CO$_2$ emissions relative to the three-sectors-aggregated emissions in 1990.} \label{fig_PHS_revenues}
\end{figure}

For most of the storage technologies the capacity and dispatch patterns are jointly optimised, and since the model assumes long-term market equilibrium, the solution implies that the sum of the costs incurred by those technologies are exactly compensated by the market revenues that they get. This is not the case for PHS whose capacities are considered exogenous, because their potential expansion in Europe is limited and assumed to be fully amortised. The PHS dispatch time series are optimised to ease the system operation. Figure \ref{fig_PHS_revenues} depicts the annual market revenues of PHS as a function of CO$_2$ emissions. The revenues $r_{PHS}$ are calculated as

\begin{equation}
	r_{PHS}=\sum_{i,t} g_{i,PHS,t} \cdot p_{i,t}
\end{equation}
where $g_{i,PHS,t}$ is the power dispatch at every country $i$ and hour $t$, and $p_{i,t}$ is the nodal price. $g_{i,PHS,t}$ is positive if PHS discharges energy and negative if it charges. 
As a general trend, PHS revenues increase as CO$_2$ emissions curb, but when the scenario include other sectors, the market revenues are affected. Figure \ref{fig_PHS_revenues} also depicts the annual market revenues of reservoir hydropower plants whose capacities are exogenous to the model and considered fully amortised too. For reservoir hydro, market revenues remains roughly constant up to a certain CO$_2$ emissions limit, which depends on the included sectors. For lower CO$_2$ emissions, revenues increases due to an increase in average nodal prices as the system becomes more expensive.

\subsection{Sensitivity to model assumptions} \label{sec_sensitivity}

\begin{figure*}[!h]
	\centering
	\includegraphics[width=0.7\textwidth]{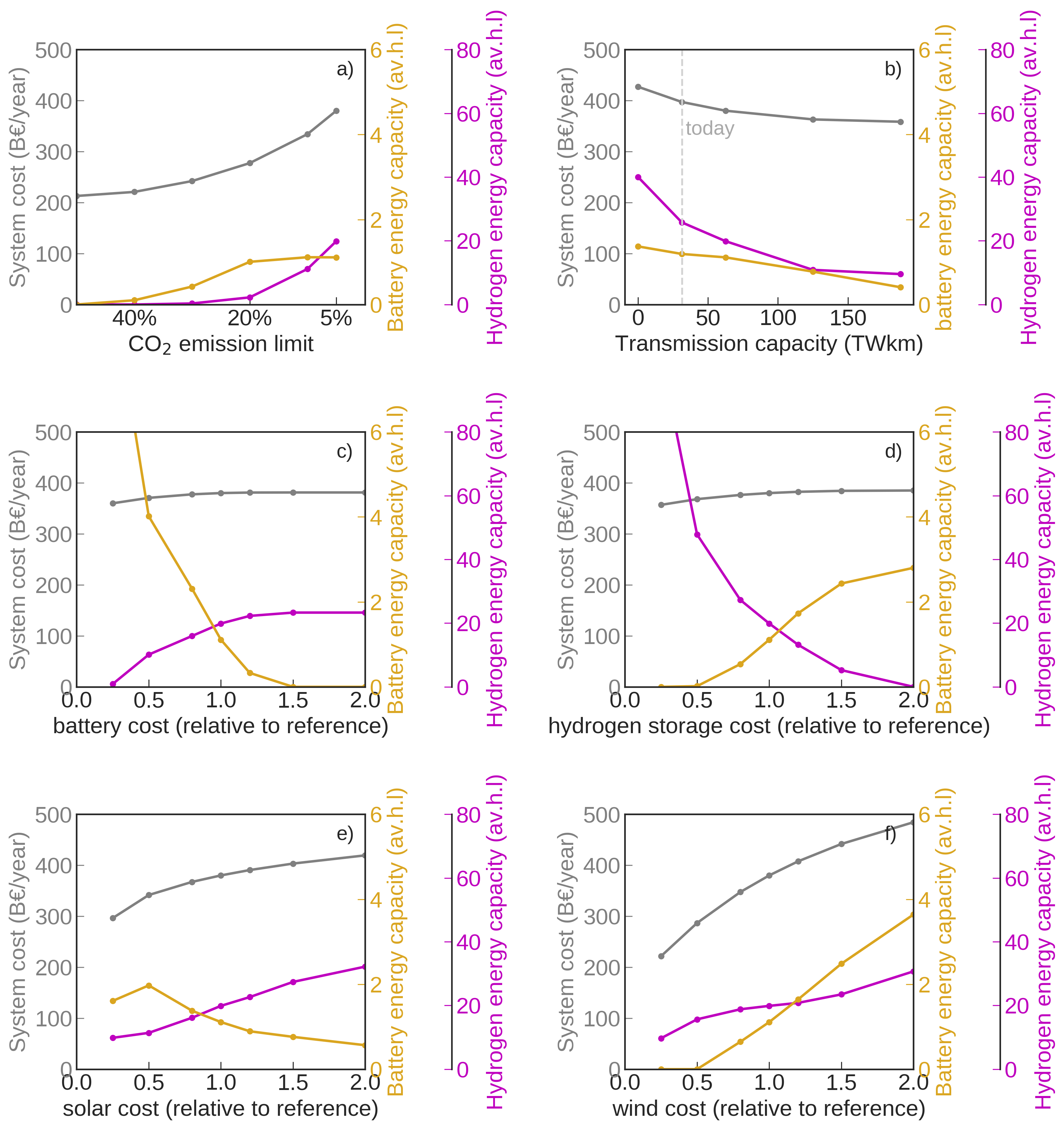}\\
	\caption{Sensitivity of storage energy capacities and system cost in the scenario Electricity+Heating to the main assumptions in the model: (a) CO$_2$ emission limit, (b) transmission expansion, (c) battery costs, (d) hydrogen storage cost, (e) solar PV cost, (f) onshore and offshore wind cost. }
	\label{fig_sensitivity}
\end{figure*}

In this section the sensitivity of the cost-effective storage energy capacities in the scenario Electricity+Heating is analysed by independently modifying some of the key assumptions of the model. We define a reference scenario where the expansion of interconnection capacity is capped to twice today's volume and maximum CO$_2$ emissions are limited to 5\% of emissions from the electricity and heating sectors in 1990. Figure \ref{fig_sensitivity}(a) shows the optimal storage capacities as a function of CO$_2$ emissions cap. Figure \ref{fig_sensitivity}(a) is the same as Figure \ref{fig_capacities} but it is replotted here to provide a reference for the other sensitivity plots. Figure \ref{fig_sensitivity}(b) shows the cost-effective energy capacities when the CO$_2$ is limited to 5\% and the transmission expansion cap is varied, Eq. (\ref{eq_cap}). When cost-optimal expansion is allowed, the system installs 159 TWkm which represents 5 times the current transmission capacity. As shown in \cite{Schlachtberger_2017, Brown_2018} for the 30 nodes network, most of the cost reduction is obtained by a compromised grid expansion that represents 3 times today's interconnection capacity. Nevertheless, we should remark here that this result is obtained when modelling only cross-border transmission. For the 256 nodes network used in \cite{Horsch_2017}, it was shown that capacity expansion equivalent to 1.25 today's value is enough to lock in most of the cost benefits of grid expansion. Reducing transmission to zero multiplies by two the cost-effective hydrogen storage because wind fluctuations in the synoptic time scale can no longer be spatially smoothed by compensating deficits and surplus of energy among countries. However, in as much as solar generation is highly correlated among countries, decreasing interconnection capacities has a lower impact on the required batteries capacity. \\

Plots (c)-(f) in Figure \ref{fig_sensitivity} depict the changes compared to the reference scenario (transmission capped to twice today's values and CO$_2$ emissions limited to 5\% of 1990 values) when the costs assumed for batteries, hydrogen storage, solar PV, and wind are varied. For batteries and hydrogen storage, the variations in cost are assumed to happen simultaneously in the cost of the energy and power capacity. It is interesting to realise that, if batteries achieve a cost half of the reference (Table \ref{tab_stores_characteristics}), it becomes competitive with hydrogen, and the required energy capacity of the latter is reduced. Conversely, if the cost of batteries does not go below 1.5 times the reference cost, it does not emerge in the system. A similar trend is observed for hydrogen storage. Reducing its reference cost (Table \ref{tab_stores_characteristics}) by half allows it to take up the role of batteries. Achieving only two times the reference cost pushes hydrogen storage out of the optimal configuration. Since the costs of both technologies are expected to remain in the vicinity of the assumed reference values, the system could benefit from different technologies to provide short and long-term storage. For hydrogen storage, underground salt cavern can reduce the cost of storage energy capacity by two orders of magnitude \cite{Steward_2009}. This would increase the competitiveness of hydrogen storage but the effect would be less pronounced than the one shown in Figure \ref{fig_sensitivity}(d) where energy and power capacity are assumed to change simultaneously. \\

Since solar PV pairs with batteries and wind with hydrogen storage, variations in the cost of generating technologies impact the cost-effective energy capacity of the storage technologies. Lower cost for solar PV favours batteries in detriment of hydrogen storage, but when solar PV cost gets reduced to 25\% of reference values (Table \ref{tab_costs}), the optimum batteries capacity decreases again. If the capital cost of PV is significantly reduced, large curtailment is allowed since its impact on the system cost is diminished. Lower cost for wind favours this source in detriment of solar PV and, consequently, the need for batteries to smooth the daily cycle decreases. The need for hydrogen is also reduced due to the lower economic impact of wind curtailment. The reader is referred to \cite{Schlachtberger_2018} for extended analyses on the influence of weather data, cost parameters and policy constraints on the power system optimal configuration.

\subsection{Comparison to similar studies} \label{comparison}

In the past, a cusp singularity for the required storage capacity has been identified when the average renewable generation equals the average electricity demand \cite{Rasmussen_2012, Jensen_2014}. Rasmussen \textit{et al.} \cite{Rasmussen_2012} calculated the necessary storage capacity to minimise the backup energy needed to balance VRES generation by using a simple dispatch algorithm for stores that neglects any consideration regarding economic cost. Put it simply, any energy deficit is covered with storage unless it runs empty and any excess generation is stored unless the store gets full. This strategy named as `storage-first' is not only simple but it is one of the optimal strategies with respect to minimising backup energy \cite{Rasmussen_2012}. Jensen and Greiner used the same dispatch strategy and showed that, assuming the wind to solar ratio that removes seasonal variation in the VRES generation time series, the strong increase in the required storage energy capacity is caused by temporal correlations on the synoptic weather time scale \cite{Jensen_2014}. In their work, the necessary storage energy capacity increases up to three orders of magnitude when the average renewable generation rises from 0.4 to 1 av.h.l.$_{el}$. The approach followed in this work is radically different. The optimal system configuration is determined for decreasing CO$_2$ emissions limit, and consequently increasing renewable penetration. In this case, the cost minimisation restrains the installation of excessive storage capacities and search for the optimal combination of storage and CO$_2$-emitting backup energy. Then, as CO$_2$ cap approaches zero, the amount of backup energy that can be used is reduced, and the system requires larger storage energy capacity showing a new diverging behaviour. In this case not when the average renewable generation approaches the average demand but when allowed CO$_2$ emissions approach zero.\\

For 5\% CO$_2$ emissions, the cost-optimal system requires total storage energy capacity equal to roughly 22 av.h.l$_{el}$ (0.87 corresponding to PHS, 1.4 to batteries and 19.4 to hydrogen storage). Due to the diverging behaviour of storage at this CO$_2$ emissions cap, the quantitative comparison with previous works is difficult. Nevertheless, a robust result of this analysis indicates that, regardless of the sectors included in the scenario, 20\% CO$_2$ emissions represents a critical target from which the system becomes more expensive and storage requirement diverges. This critical CO$_2$ reduction was already identified in \cite{Brown_2019} where expensive technologies such as methanation emerge in the European sector-coupled energy system for CO$_2$ emissions targets lower than 20\%. \\

For the European power system with renewable penetration of 89\% and moderate grid expansion, Cebulla \textit{et al.} \cite{Cebulla_2017} determined an optimal storage energy capacity of 30 TWh (approximately equivalent to 90 av.h.l$_{el}$) by using a joint capacity and dispatch optimisation model (REMIx). This storage capacity is almost entirely in the shape of hydrogen and the large capacity, compared to our result, can be explained by the fact that Cebulla assumes lower cost for hydrogen energy capacity, 1 \euro/kWh, corresponding to salt caverns. For the power system in the EU-MENA region, Bussar \textit{et al.} \cite{Bussar_2014} estimated a required storage capacity equivalent to 22 days of av.h.l$_{el}$. This extraordinary high value is obtained because the rule-based dispatch of the system components assumed in Bussar's work do not optimise the storage dispatch and, most importantly, because no backup energy is included in their model. For interconnected sector-coupled European energy system, Child \textit{et al.} \cite{Child_2018, Child_2019} identified an optimum storage capacity of 2500 GWh (approximately equivalent to 8 av.h.l$_{el}$) composed by central batteries (30\%), prosumers batteries (56\%), PHS (12\%), and CAES (2\%). Their results are not directly comparable with ours because of the slightly different regional scope of Child's analysis, the additional technologies considered by the authors, \textit{i.e.} power to gas, and the independent modelling of prosumers which emphasises the role of home batteries associated with PV installations. Nevertheless, Child \textit{et al.} coincide in identifying 80\% as the renewable penetration from which large storage energy capacities become cost effective.  Cebulla \textit{et al.} \cite{Cebulla_2018} plotted the optimal electricity storage estimated by different authors and observed that storage power capacity raises linearly to renewable penetration while storage energy capacity increases exponentially. The authors also identified as outliers large storage capacities obtained when high CO$_2$ price is assumed. This anticipates the results that we have obtained, that is, not only renewable penetration impacts the storage requirement, but restraining the available backup energy triggers the required energy and power storage capacities. The available backup energy can be restrained by a certain CO$_2$ emissions constraint, as we do here, or by assuming a high CO$_2$ tax. Conversely, including in the model the possibility of methanation releases the available backup energy and consequently reduces the required storage capacities. \\

The strong link identified between solar generation-batteries dispatch and wind generation-hydrogen storage dispatch supports previous findings \cite{Rasmussen_2012, Schlachtberger_2017, Cebulla_2017, Brown_2018}. Recently flow-tracing techniques have been used to unveil the use of every kind of storage by different generation technologies \cite{Tranberg_2018}. Flow-tracing agrees with our analyses also when highlighting the strong link between PV generation and PHS dispatch. Concerning storage power-to-energy capacity, for the electricity sector, Schlachtberger \textit{et al.} \cite{Schlachtberger_2017} assumed a fixed 6 hours ratio for batteries, which is very similar to the 5.6 hours that we have obtained by independently optimising power and energy capacity. For hydrogen storage, they assumed one week as discharge time while our independent optimisation showed roughly two days. Finally, it is worth mentioning that the system cost predicted here is slightly lower than that estimated in \cite{Brown_2018}. The reason is that, based on the Danish Energy Technology Catalogue \cite{DEA_2016}, more aggressive cost assumptions have been made for onshore wind, both on the investment cost and the technical lifetime.  

\subsection{Limitations of the analysis}

The analysis performed here entails some limitations. First, only electricity, heating, and transport sectors are included but decarbonising the remaining sectors (industry, shipping, aviation or agriculture) might be even more challenging. Furthermore, several possible links among sectors are missing such as the possibility of using the hydrogen produced via renewable surplus and electrolysis to power the shipping sector, or producing synthetic methane using hydrogen and CO$_2$. This last possibility was modelled in \cite{Brown_2018, Brown_2019} and might impact the results as discussed in Section \ref{comparison}. The lacking of biomass in the model influences the high CO$_2$ shadow price found here, as discussed in Section \ref{sec_capacities_evolution}. Nuclear, coal, and Carbon Capture and Storage (CCS) are not included in the model. Producing hydrogen with steam reforming of methane and sequestering the CO$_2$ underground could be cheaper than electrolysed-produced hydrogen. This could alter the role of hydrogen and its cost-effective storage capacity. Second, the model includes electric batteries, hydrogen storage, PHS, individual and central thermal energy storage but other storage technologies are not modelled, see Section \ref{sec_storage_technologies}. Third, a greenfield optimisation is performed, but the current generation capacity layout in Europe might have a considerable influence in the configuration of the future European energy system. Fourth, the optimal configuration is impacted by cost assumptions and imposed constraints. For the most critical parameters, their effect was analysed in Section \ref{sec_sensitivity}, but changes in other costs assumptions may also impact the results. Fifth, only one year has been used to represent variability of renewable generation. In \cite{Schlachtberger_2018}, 4 years of weather data were used to analyse the system sensitivity and the  cost of the power sector was found to be stable, although extensive sensitivity analysis to weather data such the one in \cite{Collins_2018} are desirable. Sixth, the coarse-grained network used here might also impact the results. For the power sector, H{\"o}rsch and Brown \cite{Horsch_2017} found that the system cost remains roughly constant due to the counterbalancing of two effects as a higher number of nodes are used to represent the network: sites with high capacity factors for VRES are available for a more finely resolved network, but the emergence of bottlenecks inside countries prevent the use of wind generated at exterior nodes with high capacity factors.  Seventh, the role of storage technologies has been evaluated here in terms of energy arbitrage, that is, shifting energy from low-price periods with excess of VRES production to high-price periods where generation is insufficient. However, some storage technologies may become more competitive when also taking into account the provision of other services to the system such as power reliability, secondary or tertiary response, black start of the system or maximisation of local self-consumption \cite{Schmidt_2019}.  Finally, regarding the assumptions in the transport sector it should be remarked that a significant reduction in the number of cars in Europe will be essential to achieve the necessary decarbonisation in this sector. Increasing the EV utilisation through car-sharing, decreasing the size and weight of vehicles, and increasing the number of passengers per car are some of the necessary strategies to curb the CO$_2$ emissions associated not only with the use of EVs but also with manufacturing \cite{circular_economy}. From the system point of view, as discussed in \cite{Brown_2018}, assuming lower availability of BEV has a little impact on the results since the total charging power, 1.3 TW, is much higher than the average hourly demand from transport, 125 GWh. 

\section{Conclusions} \label{sec_conclusions}

In this paper, we have investigated the role of different storage technologies, under variable CO$_2$ emissions caps and sector-coupled scenarios, by modelling Europe as a one-node-per-country network and by using economic optimisation. The capacities for generation and storage technologies are optimised together with the hourly dispatch for a full year. An extension of transmission capacity volume up to twice of today's value is allowed and renewable generation in every country is assumed to be proportional to the electricity demand.\\

As the electricity-related CO$_2$ emissions are restricted, the need for storage does not increase linearly but it remains almost negligible for low reductions and ramps up for CO$_2$ reductions higher than 80\%, relative to 1990. Because synthetic production of methane is not included in the model, a strict CO$_2$ emission cap limits the available backup energy and requires large storage energy capacities to counterbalance renewable fluctuations. For 5\% CO$_2$ emissions, we found that batteries and hydrogen storage optimal energy capacity is respectively 455 GWh (equivalent to 1.4 average electricity demand) and 6.3 TWh (19.4 av.h.l$_{el}$). Both values are well below the estimated technical potentials. On the one hand, batteries have higher efficiency and higher unitary cost for energy capacity, so it is cost optimal as short-term storage to counterbalance solar PV generation. On the other hand, hydrogen storage, with lower round-trip efficiency, but also lower unitary cost for energy capacity, is preferred to smooth multi-day wind fluctuations. Consequently, the coexistence of two different storage technologies to arbitrage energy at different time scale is expected in the future European power system. This result is found to be notably robust. The cannibalisation of both time-scale storage requirements by one single technology, either batteries or hydrogen storage, only happens when the cost assumptions are varied by more than $\pm$ 50\%.\\

One of the main benefits of sector coupling is that higher global CO$_2$ reductions can be achieved before large storage capacities emerge. This is a strong message for policymakers, \textit{i.e.}, in the transition to a low-carbon European energy system, coupling the sectors delays the moment in which large storage capacities are needed and allows time to further develop storage technologies. The results also showed that, if half of the current European car fleet is transformed to electric vehicles that can feed back into the grid, the need for stationary batteries is completely avoided. Coupling the heating sector brings large capacities of centralised thermal energy storage to smooth the significant seasonal variation of heat demand. Neither coupling the transport nor the heating sector decreases the need for hydrogen storage (or an alternative technology with low energy capacity unitary cost) but sector coupling allows further CO$_2$ reductions before this capacity is needed. \\

Moreover, we found that the optimal dispatch patterns of different storage technologies are determined by two factors: the CO$_2$ emissions cap and the additional available technologies. This is particularly relevant for Pumped Hydro Storage (PHS) and reservoir hydro whose capacities are not expected to increase in Europe. For instance, for the power system, when CO$_2$ emissions are restricted from 60\% to 5\%, PHS considerably increases its cycling frequency, raising its global market revenues by 35\%. However, coupling the power system with other sectors expands the demand, which reduces PHS cycling frequency, and brings additional storage to the system, \textit{e.g}. EV batteries, which dramatically reduces PHS revenues. \\

The highly-renewable highly-interconnected future European energy system will be more complex to design and operate. In order to unveil the dependencies between the generation, storage, and interconnection elements in the system, models with sufficient temporal, spatial and technological resolution, and appropriate representation of sector-coupling and grid interconnections, such as the one applied in this paper, are needed.

\section{Acknowledgements}
M. Victoria, K. Zhu, G. B. Andresen and M. Greiner are fully or partially funded by the RE-INVEST project, which is supported by  the  Innovation  Fund  Denmark  under  grant  number  6154-00022B. T.B. acknowledges funding from the Helmholtz Association under grant no. VH-NG-1352. The responsibility for the contents lies solely with the authors.

\section{References}

\bibliography{energy}

\begin{thebibliography}{10}
\expandafter\ifx\csname url\endcsname\relax
  \def\url#1{\texttt{#1}}\fi
\expandafter\ifx\csname urlprefix\endcsname\relax\def\urlprefix{URL }\fi
\expandafter\ifx\csname href\endcsname\relax
  \def\href#1#2{#2} \def\path#1{#1}\fi

\bibitem{IPCC_1.5}
\href{https://www.ipcc.ch/sr15/}{Global {Warming} of 1.5$^{\circ}${C}, an
  {IPCC} special report on the impacts of global warming of 1.5$^{\circ}${C}
  above preindustrial levels and related global greenhouse gas emission
  pathways, in the context of strengthening the global response to the threat
  of climate change, sustainable development, and efforts to eradicate poverty,
  {Intergovernmental} {Panel} on {Climate} {Change} ({IPCC})}, Tech. rep.
  (2018).
\newline\urlprefix\url{https://www.ipcc.ch/sr15/}

\bibitem{in-depth_2018}
\href{https://ec.europa.eu/clima/news/commission-calls-climate-neutral-europe-2050_en}{In-depth
  analysis in support of the {Comission} {Communication} {COM}(2018) 773 {A}
  {Clean} {Planet} for all. {A} {European} long-term strategic vision for a
  prosperous, modern, competitive and climate neutral economy}, Tech. rep.
  (Nov. 2018).
\newline\urlprefix\url{https://ec.europa.eu/clima/news/commission-calls-climate-neutral-europe-2050_en}

\bibitem{Heide_2011}
D.~Heide, M.~Greiner, L.~von Bremen, C.~Hoffmann,
  \href{http://www.sciencedirect.com/science/article/pii/S0960148111000851}{Reduced
  storage and balancing needs in a fully renewable {European} power system with
  excess wind and solar power generation}, Renewable Energy 36~(9) (2011)
  2515--2523.
\newblock \href {http://dx.doi.org/10.1016/j.renene.2011.02.009}
  {\path{doi:10.1016/j.renene.2011.02.009}}.
\newline\urlprefix\url{http://www.sciencedirect.com/science/article/pii/S0960148111000851}

\bibitem{Rasmussen_2012}
M.~G. Rasmussen, G.~B. Andresen, M.~Greiner,
  \href{http://www.sciencedirect.com/science/article/pii/S0301421512007677}{Storage
  and balancing synergies in a fully or highly renewable pan-{European} power
  system}, Energy Policy 51 (2012) 642 -- 651.
\newblock \href {http://dx.doi.org/https://doi.org/10.1016/j.enpol.2012.09.009}
  {\path{doi:https://doi.org/10.1016/j.enpol.2012.09.009}}.
\newline\urlprefix\url{http://www.sciencedirect.com/science/article/pii/S0301421512007677}

\bibitem{Jensen_2014}
T.~V. Jensen, M.~Greiner,
  \href{https://link.springer.com/article/10.1140/epjst/e2014-02216-9}{Emergence
  of a phase transition for the required amount of storage in highly renewable
  electricity systems}, The European Physical Journal Special Topics 223~(12)
  (2014) 2475--2481.
\newblock \href {http://dx.doi.org/10.1140/epjst/e2014-02216-9}
  {\path{doi:10.1140/epjst/e2014-02216-9}}.
\newline\urlprefix\url{https://link.springer.com/article/10.1140/epjst/e2014-02216-9}

\bibitem{Andresen_2014}
G.~B. Andresen, R.~A. Rodriguez, S.~Becker, M.~Greiner,
  \href{http://www.sciencedirect.com/science/article/pii/S0360544214002977}{The
  potential for arbitrage of wind and solar surplus power in {Denmark}}, Energy
  76 (2014) 49--58.
\newblock \href {http://dx.doi.org/10.1016/j.energy.2014.03.033}
  {\path{doi:10.1016/j.energy.2014.03.033}}.
\newline\urlprefix\url{http://www.sciencedirect.com/science/article/pii/S0360544214002977}

\bibitem{Bussar_2014}
C.~Bussar, M.~Moos, R.~Alvarez, P.~Wolf, T.~Thien, H.~Chen, Z.~Cai,
  M.~Leuthold, D.~U. Sauer, A.~Moser,
  \href{http://www.sciencedirect.com/science/article/pii/S1876610214001726}{Optimal
  {Allocation} and {Capacity} of {Energy} {Storage} {Systems} in a {Future}
  {European} {Power} {System} with 100\% {Renewable} {Energy} {Generation}},
  Energy Procedia 46 (2014) 40--47.
\newblock \href {http://dx.doi.org/10.1016/j.egypro.2014.01.156}
  {\path{doi:10.1016/j.egypro.2014.01.156}}.
\newline\urlprefix\url{http://www.sciencedirect.com/science/article/pii/S1876610214001726}

\bibitem{Bussar_2016}
C.~Bussar, P.~Stöcker, Z.~Cai, L.~Moraes~Jr., D.~Magnor, P.~Wiernes, N.~v.
  Bracht, A.~Moser, D.~U. Sauer,
  \href{http://www.sciencedirect.com/science/article/pii/S2352152X16300135}{Large-scale
  integration of renewable energies and impact on storage demand in a
  {European} renewable power system of 2050—{Sensitivity} study}, Journal of
  Energy Storage 6 (2016) 1--10.
\newblock \href {http://dx.doi.org/10.1016/j.est.2016.02.004}
  {\path{doi:10.1016/j.est.2016.02.004}}.
\newline\urlprefix\url{http://www.sciencedirect.com/science/article/pii/S2352152X16300135}

\bibitem{Schlachtberger_2017}
D.~P. Schlachtberger, T.~Brown, S.~Schramm, M.~Greiner,
  \href{http://www.sciencedirect.com/science/article/pii/S0360544217309969}{The
  benefits of cooperation in a highly renewable {European} electricity
  network}, Energy 134~(Supplement C) (2017) 469--481.
\newblock \href {http://dx.doi.org/10.1016/j.energy.2017.06.004}
  {\path{doi:10.1016/j.energy.2017.06.004}}.
\newline\urlprefix\url{http://www.sciencedirect.com/science/article/pii/S0360544217309969}

\bibitem{Cebulla_2017}
F.~Cebulla, T.~Naegler, M.~Pohl,
  \href{http://www.sciencedirect.com/science/article/pii/S2352152X17302815}{Electrical
  energy storage in highly renewable {European} energy systems: {Capacity}
  requirements, spatial distribution, and storage dispatch}, Journal of Energy
  Storage 14 (2017) 211--223.
\newblock \href {http://dx.doi.org/10.1016/j.est.2017.10.004}
  {\path{doi:10.1016/j.est.2017.10.004}}.
\newline\urlprefix\url{http://www.sciencedirect.com/science/article/pii/S2352152X17302815}

\bibitem{Cebulla_2018}
F.~Cebulla, J.~Haas, J.~Eichman, W.~Nowak, P.~Mancarella,
  \href{http://www.sciencedirect.com/science/article/pii/S0959652618301665}{How
  much electrical energy storage do we need? {A} synthesis for the {U.S.},
  {Europe}, and {Germany}} 181  449--459.
\newblock \href {http://dx.doi.org/10.1016/j.jclepro.2018.01.144}
  {\path{doi:10.1016/j.jclepro.2018.01.144}}.
\newline\urlprefix\url{http://www.sciencedirect.com/science/article/pii/S0959652618301665}

\bibitem{Rodriguez_2014}
R.~A. Rodr\'iguez, S.~Becker, G.~B. Andresen, D.~Heide, M.~Greiner,
  \href{http://www.sciencedirect.com/science/article/pii/S0960148113005351}{Transmission
  needs across a fully renewable {European} power system}, Renewable Energy 63
  (2014) 467--476.
\newblock \href {http://dx.doi.org/10.1016/j.renene.2013.10.005}
  {\path{doi:10.1016/j.renene.2013.10.005}}.
\newline\urlprefix\url{http://www.sciencedirect.com/science/article/pii/S0960148113005351}

\bibitem{Eriksen_2017}
E.~H. Eriksen, L.~J. Schwenk-Nebbe, B.~Tranberg, T.~Brown, M.~Greiner,
  \href{http://www.sciencedirect.com/science/article/pii/S0360544217309593}{Optimal
  heterogeneity in a simplified highly renewable {European} electricity
  system}, Energy 133~(Supplement C) (2017) 913--928.
\newblock \href {http://dx.doi.org/10.1016/j.energy.2017.05.170}
  {\path{doi:10.1016/j.energy.2017.05.170}}.
\newline\urlprefix\url{http://www.sciencedirect.com/science/article/pii/S0360544217309593}

\bibitem{Gils_2017a}
H.~C. Gils, Y.~Scholz, T.~Pregger, D.~L. de~Tena, D.~Heide,
  \href{http://www.sciencedirect.com/science/article/pii/S0360544217301238}{Integrated
  modelling of variable renewable energy-based power supply in {Europe}},
  Energy 123 (2017) 173 -- 188.
\newblock \href
  {http://dx.doi.org/https://doi.org/10.1016/j.energy.2017.01.115}
  {\path{doi:https://doi.org/10.1016/j.energy.2017.01.115}}.
\newline\urlprefix\url{http://www.sciencedirect.com/science/article/pii/S0360544217301238}

\bibitem{Sorensen_1975}
B.~S{\o}rensen,
  \href{https://science.sciencemag.org/content/189/4199/255}{Energy and
  {Resources}: {A} plan is outlined according to which solar and wind energy
  would supply {Denmark}'s needs by the year 2050}, Science 189~(4199) (1975)
  255--260.
\newblock \href {http://dx.doi.org/10.1126/science.189.4199.255}
  {\path{doi:10.1126/science.189.4199.255}}.
\newline\urlprefix\url{https://science.sciencemag.org/content/189/4199/255}

\bibitem{Connolly_2016}
D.~Connolly, H.~Lund, B.~V. Mathiesen,
  \href{http://www.sciencedirect.com/science/article/pii/S1364032116002331}{Smart
  {Energy} {Europe}: {The} technical and economic impact of one potential 100\%
  renewable energy scenario for the {European} {Union}}, Renewable and
  Sustainable Energy Reviews 60 (2016) 1634--1653.
\newblock \href {http://dx.doi.org/10.1016/j.rser.2016.02.025}
  {\path{doi:10.1016/j.rser.2016.02.025}}.
\newline\urlprefix\url{http://www.sciencedirect.com/science/article/pii/S1364032116002331}

\bibitem{Brown_2018}
T.~Brown, D.~Schlachtberger, A.~Kies, S.~Schramm, M.~Greiner,
  \href{http://www.sciencedirect.com/science/article/pii/S036054421831288X}{Synergies
  of sector coupling and transmission reinforcement in a cost-optimised, highly
  renewable {European} energy system}, Energy 160 (2018) 720--739.
\newblock \href {http://dx.doi.org/10.1016/j.energy.2018.06.222}
  {\path{doi:10.1016/j.energy.2018.06.222}}.
\newline\urlprefix\url{http://www.sciencedirect.com/science/article/pii/S036054421831288X}

\bibitem{Brown_2019}
T.~Brown, M.~Sch{\"a}fer, M.~Greiner,
  \href{https://www.mdpi.com/1996-1073/12/6/1032}{Sectoral {Interactions} as
  {Carbon} {Dioxide} {Emissions} {Approach} {Zero} in a {Highly}-{Renewable}
  {European} {Energy} {System}}, Energies 12~(6) (2019) 1032.
\newblock \href {http://dx.doi.org/10.3390/en12061032}
  {\path{doi:10.3390/en12061032}}.
\newline\urlprefix\url{https://www.mdpi.com/1996-1073/12/6/1032}

\bibitem{Zhu_2019}
K.~Zhu, M.~Victoria, T.~Brown, G.~B. Andresen, M.~Greiner,
  \href{http://www.sciencedirect.com/science/article/pii/S030626191831835X}{Impact
  of {CO}2 prices on the design of a highly decarbonised coupled electricity
  and heating system in {E}urope} 236 (2019) 622--634.
\newblock \href {http://dx.doi.org/10.1016/j.apenergy.2018.12.016}
  {\path{doi:10.1016/j.apenergy.2018.12.016}}.
\newline\urlprefix\url{http://www.sciencedirect.com/science/article/pii/S030626191831835X}

\bibitem{Child_2018}
M.~Child, D.~Bogdanov, C.~Breyer,
  \href{http://www.sciencedirect.com/science/article/pii/S1876610218310221}{The
  role of storage technologies for the transition to a 100\% renewable energy
  system in {Europe}}, Energy Procedia 155 (2018) 44--60.
\newblock \href {http://dx.doi.org/10.1016/j.egypro.2018.11.067}
  {\path{doi:10.1016/j.egypro.2018.11.067}}.
\newline\urlprefix\url{http://www.sciencedirect.com/science/article/pii/S1876610218310221}

\bibitem{Child_2019}
M.~Child, C.~Kemfert, D.~Bogdanov, C.~Breyer,
  \href{http://www.sciencedirect.com/science/article/pii/S0960148119302319}{Flexible
  electricity generation, grid exchange and storage for the transition to a
  100\% renewable energy system in {Europe}}, Renewable Energy 139 (2019)
  80--101.
\newblock \href {http://dx.doi.org/10.1016/j.renene.2019.02.077}
  {\path{doi:10.1016/j.renene.2019.02.077}}.
\newline\urlprefix\url{http://www.sciencedirect.com/science/article/pii/S0960148119302319}

\bibitem{Schroeder_2013}
A.~Schr\"oeder, F.~Kunz, F.~Meiss, R.~Mendelevitch, C.~von Hirschhausen,
  \href{https://www.econstor.eu/handle/10419/80348}{Current and prospective
  costs of electricity generation until 2050, {Data} {Documentation}, {DIW} 68.
  {Berlin}: {Deutsches} {Institut}}.
\newline\urlprefix\url{https://www.econstor.eu/handle/10419/80348}

\bibitem{Budischak_2013}
C.~Budischak, D.~Sewell, H.~Thomson, L.~Mach, D.~E. Veron, W.~Kempton,
  \href{http://www.sciencedirect.com/science/article/pii/S0378775312014759}{Cost-minimized
  combinations of wind power, solar power and electrochemical storage, powering
  the grid up to 99.9\% of the time}, Journal of Power Sources 225 (2013)
  60--74.
\newblock \href {http://dx.doi.org/10.1016/j.jpowsour.2012.09.054}
  {\path{doi:10.1016/j.jpowsour.2012.09.054}}.
\newline\urlprefix\url{http://www.sciencedirect.com/science/article/pii/S0378775312014759}

\bibitem{Steward_2009}
D.~Steward, G.~Saur, M.~Penev, T.~Ramsden,
  \href{https://www.nrel.gov/docs/fy10osti/46719.pdf}{Lifecycle cost analysis
  of hydrogen versus other technologies for electrical energy storage {NREL}},
  Tech. rep. (2009).
\newline\urlprefix\url{https://www.nrel.gov/docs/fy10osti/46719.pdf}

\bibitem{Henning_2014}
H.-M. Henning, A.~Palzer,
  \href{http://www.sciencedirect.com/science/article/pii/S1364032113006710}{A
  comprehensive model for the {German} electricity and heat sector in a future
  energy system with a dominant contribution from renewable energy
  technologies—{Part} {I}: {Methodology}}, Renewable and Sustainable Energy
  Reviews 30 (2014) 1003--1018.
\newblock \href {http://dx.doi.org/10.1016/j.rser.2013.09.012}
  {\path{doi:10.1016/j.rser.2013.09.012}}.
\newline\urlprefix\url{http://www.sciencedirect.com/science/article/pii/S1364032113006710}

\bibitem{Lund_2016}
H.~Lund, P.~A. {\O}stergaard, D.~Connolly, I.~Ridjan, B.~V. Mathiesen,
  F.~Hvelplund, J.~Z. Thellufsen, P.~Sorknæs,
  \href{https://journals.aau.dk/index.php/sepm/article/view/1574}{Energy
  {Storage} and {Smart} {Energy} {Systems}}, International Journal of
  Sustainable Energy Planning and Management 11 (2016) 3--14.
\newblock \href {http://dx.doi.org/10.5278/ijsepm.2016.11.2}
  {\path{doi:10.5278/ijsepm.2016.11.2}}.
\newline\urlprefix\url{https://journals.aau.dk/index.php/sepm/article/view/1574}

\bibitem{Schmidt_2017}
O.~Schmidt, A.~Hawkes, A.~Gambhir, I.~Staffell,
  \href{https://www.nature.com/articles/nenergy2017110}{The future cost of
  electrical energy storage based on experience rates}, Nature Energy 2~(8)
  (2017) 17110.
\newblock \href {http://dx.doi.org/10.1038/nenergy.2017.110}
  {\path{doi:10.1038/nenergy.2017.110}}.
\newline\urlprefix\url{https://www.nature.com/articles/nenergy2017110}

\bibitem{Kittner_2017}
N.~Kittner, F.~Lill, D.~M. Kammen,
  \href{https://www.nature.com/articles/nenergy2017125}{Energy storage
  deployment and innovation for the clean energy transition}, Nature Energy
  2~(9) (2017) 17125.
\newblock \href {http://dx.doi.org/10.1038/nenergy.2017.125}
  {\path{doi:10.1038/nenergy.2017.125}}.
\newline\urlprefix\url{https://www.nature.com/articles/nenergy2017125}

\bibitem{recycle_batteries}
\href{https://www.nature.com/articles/s41560-019-0376-4}{Recycle spent
  batteries}, Nature Energy 4 (2019) 253.
\newblock \href {http://dx.doi.org/10.1038/s41560-019-0376-4}
  {\path{doi:10.1038/s41560-019-0376-4}}.
\newline\urlprefix\url{https://www.nature.com/articles/s41560-019-0376-4}

\bibitem{Staffell_2019}
I.~Staffell, D.~Scamman, A.~V. Abad, P.~Balcombe, P.~E. Dodds, P.~Ekins,
  N.~Shah, K.~R. Ward,
  \href{https://pubs.rsc.org/en/content/articlelanding/2019/ee/c8ee01157e}{The
  role of hydrogen and fuel cells in the global energy system}, Energy \&
  Environmental Science 12~(2) (2019) 463--491.
\newblock \href {http://dx.doi.org/10.1039/C8EE01157E}
  {\path{doi:10.1039/C8EE01157E}}.
\newline\urlprefix\url{https://pubs.rsc.org/en/content/articlelanding/2019/ee/c8ee01157e}

\bibitem{Gilhaus_2007}
A.~Gilhaus, Natural {Gas} {Storage} in {Salt} {Caverns} - {Present} {Status},
  {Developments} and {Future} {Trends} in {Europe}, {Technical} {Conference}
  {Paper}, {Solution} {Mining} {Research} {Institute}, Tech. rep. (2007).

\bibitem{sunstorage}
\href{https://www.underground-sun-storage.at/en/project/project-description.html}{Underground
  {Sun} {Storage} - {Project} webpage}.
\newline\urlprefix\url{https://www.underground-sun-storage.at/en/project/project-description.html}

\bibitem{Julch_2016}
V.~J{\"u}lch,
  \href{http://www.sciencedirect.com/science/article/pii/S0306261916312740}{Comparison
  of electricity storage options using levelized cost of storage ({LCOS})
  method} 183  1594--1606.
\newblock \href {http://dx.doi.org/10.1016/j.apenergy.2016.08.165}
  {\path{doi:10.1016/j.apenergy.2016.08.165}}.
\newline\urlprefix\url{http://www.sciencedirect.com/science/article/pii/S0306261916312740}

\bibitem{Schmidt_2019}
O.~Schmidt, S.~Melchior, A.~Hawkes, I.~Staffell,
  \href{http://www.sciencedirect.com/science/article/pii/S254243511830583X}{Projecting
  the future levelized cost of electricity storage technologies} 3~(1) (2019)
  81--100.
\newblock \href {http://dx.doi.org/10.1016/j.joule.2018.12.008}
  {\path{doi:10.1016/j.joule.2018.12.008}}.
\newline\urlprefix\url{http://www.sciencedirect.com/science/article/pii/S254243511830583X}

\bibitem{Pawel_2014}
I.~Pawel,
  \href{http://www.sciencedirect.com/science/article/pii/S1876610214001751}{The
  cost of storage – how to calculate the levelized cost of stored energy
  ({LCOE}) and applications to renewable energy generation}, Energy Procedia 46
  (2014) 68--77.
\newblock \href {http://dx.doi.org/10.1016/j.egypro.2014.01.159}
  {\path{doi:10.1016/j.egypro.2014.01.159}}.
\newline\urlprefix\url{http://www.sciencedirect.com/science/article/pii/S1876610214001751}

\bibitem{circular_economy}
\href{https://www.sitra.fi/en/publications/circular-economy-powerful-force-climate-mitigation/}{The
  circular economy – {A} powerful force for climate mitigation, {Material}
  {Economics}}, Tech. rep. (2018).
\newline\urlprefix\url{https://www.sitra.fi/en/publications/circular-economy-powerful-force-climate-mitigation/}

\bibitem{PyPSA}
T.~Brown, J.~H\"orsch, D.~Schlachtberger,
  \href{https://doi.org/10.5334/jors.188}{{PyPSA: Python for Power System
  Analysis}}, Journal of Open Research Software 6.
\newblock \href {http://dx.doi.org/10.5334/jors.188}
  {\path{doi:10.5334/jors.188}}.
\newline\urlprefix\url{https://doi.org/10.5334/jors.188}

\bibitem{DEA_2016}
\href{https://ens.dk/en/our-services/projections-and-models/technology-data/technology-data-generation-electricity-and}{Technology
  {Data} for {Energy} {Plants} for {Electricity} and {District} heating
  generation}, Tech. rep., Danish Energy Agency and Energinet.dk (2016).
\newline\urlprefix\url{https://ens.dk/en/our-services/projections-and-models/technology-data/technology-data-generation-electricity-and}

\bibitem{Vartiainen_2017}
E.~Vartiainen, G.~Masson, C.~Breyer,
  \href{http://www.etip-pv.eu/fileadmin/Documents/ETIP_PV_Publications_2017-2018/LCOE_Report_March_2017.pdf}{The
  true competitiveness of solar {PV}: a {European} case study}, Tech. rep.,
  European technology and innovation platform for photovoltaics (2017).
\newline\urlprefix\url{http://www.etip-pv.eu/fileadmin/Documents/ETIP_PV_Publications_2017-2018/LCOE_Report_March_2017.pdf}

\bibitem{Palzer_thesis}
A.~Palzer, Sektor{\"u}bergreifende {Modellierung} und {Optimierung} eines
  zuk{\"u}nftigen deutschen {Energiesystems} unter {Ber{\"u}cksichtigung} von
  {Energieeffizienzma{\ss}nahmen} im {Geb{\"a}udesektor}, Ph.D. thesis, KIT
  (2016).

\bibitem{Schaber_2013}
K.~Schaber, \href{https://d-nb.info/1058680781/34}{Integration of {Variable}
  {Renewable} {Energies} in the {European} power system: a model-based analysis
  of transmission grid extensions and energy sector coupling}, Ph.D. thesis, TU
  {München} (2013).
\newline\urlprefix\url{https://d-nb.info/1058680781/34}

\bibitem{Hagspiel_2014}
S.~Hagspiel, C.~J{\"a}gemann, D.~Lindenberger, T.~Brown, S.~Cherevatskiy,
  E.~Tr{\"o}ster,
  \href{http://www.sciencedirect.com/science/article/pii/S0360544214000322}{Cost-optimal
  power system extension under flow-based market coupling}, Energy 66 (2014)
  654--666.
\newblock \href {http://dx.doi.org/10.1016/j.energy.2014.01.025}
  {\path{doi:10.1016/j.energy.2014.01.025}}.
\newline\urlprefix\url{http://www.sciencedirect.com/science/article/pii/S0360544214000322}

\bibitem{ODYSSEE}
\href{http://www.indicators.odyssee-mure.eu}{{ODYSSEE} database on energy
  effciency data \& indicators}, Tech. rep. (2016).
\newline\urlprefix\url{http://www.indicators.odyssee-mure.eu}

\bibitem{Schlachtberger_2016}
D.~P. Schlachtberger, S.~Becker, S.~Schramm, M.~Greiner,
  \href{http://www.sciencedirect.com/science/article/pii/S0196890416302606}{Backup
  flexibility classes in emerging large-scale renewable electricity systems},
  Energy Conversion and Management 125 (2016) 336--346.
\newblock \href {http://dx.doi.org/10.1016/j.enconman.2016.04.020}
  {\path{doi:10.1016/j.enconman.2016.04.020}}.
\newline\urlprefix\url{http://www.sciencedirect.com/science/article/pii/S0196890416302606}

\bibitem{Horsch_2017}
J.~H{\"o}rsch, T.~Brown, The role of spatial scale in joint optimisations of
  generation and transmission for {European} highly renewable scenarios, in:
  14th {International} {Conference} on the {European} {Energy} {Market}
  ({EEM}), 2017, pp. 1--7.
\newblock \href {http://dx.doi.org/10.1109/EEM.2017.7982024}
  {\path{doi:10.1109/EEM.2017.7982024}}.

\bibitem{Schlachtberger_2018}
D.~P. Schlachtberger, T.~Brown, M.~Schäfer, S.~Schramm, M.~Greiner,
  \href{http://www.sciencedirect.com/science/article/pii/S0360544218316025}{Cost
  optimal scenarios of a future highly renewable european electricity system:
  Exploring the influence of weather data, cost parameters and policy
  constraints}, Energy 163 (2018) 100--114.
\newblock \href {http://dx.doi.org/10.1016/j.energy.2018.08.070}
  {\path{doi:10.1016/j.energy.2018.08.070}}.
\newline\urlprefix\url{http://www.sciencedirect.com/science/article/pii/S0360544218316025}

\bibitem{Tranberg_2018}
B.~Tranberg, M.~Sch{\"a}fer, T.~Brown, J.~Hörsch, M.~Greiner, Flow-based
  analysis of storage usage in a low-carbon european electricity scenario, in:
  2018 15th International Conference on the European Energy Market ({EEM}), pp.
  1--5.
\newblock \href {http://dx.doi.org/10.1109/EEM.2018.8469951}
  {\path{doi:10.1109/EEM.2018.8469951}}.

\bibitem{Collins_2018}
S.~Collins, P.~Deane, B.~Ó~Gallachóir, S.~Pfenninger, I.~Staffell,
  \href{http://www.sciencedirect.com/science/article/pii/S254243511830285X}{Impacts
  of {Inter}-annual {Wind} and {Solar} {Variations} on the {European} {Power}
  {System}}, Joule 10 (2018) 2076--2090.
\newblock \href {http://dx.doi.org/10.1016/j.joule.2018.06.020}
  {\path{doi:10.1016/j.joule.2018.06.020}}.
\newline\urlprefix\url{http://www.sciencedirect.com/science/article/pii/S254243511830285X}

\bibitem{Brown_2016}
T.~Brown, P.~Schierhorn, E.~Tr{\"o}ster, T.~Ackermann, Optimising the european
  transmission system for 77\% renewable electricity by 2030 10~(1)  3--9.
\newblock \href {http://dx.doi.org/10.1049/iet-rpg.2015.0135}
  {\path{doi:10.1049/iet-rpg.2015.0135}}.

\bibitem{Horsch_2018}
J.~H{\"o}rsch, H.~Ronellenfitsch, D.~Witthaut, T.~Brown,
  \href{http://www.sciencedirect.com/science/article/pii/S0378779617305138}{Linear
  optimal power flow using cycle flows} 158 (2019) 126--135.
\newblock \href {http://dx.doi.org/10.1016/j.epsr.2017.12.034}
  {\path{doi:10.1016/j.epsr.2017.12.034}}.
\newline\urlprefix\url{http://www.sciencedirect.com/science/article/pii/S0378779617305138}

\bibitem{ENTSOE}
\href{https://www.entsoe.eu/data/data-portal/}{European {Transmission} {System}
  {Operators}, {Country}-specific hourly load data consumption (2011).}
\newline\urlprefix\url{https://www.entsoe.eu/data/data-portal/}

\bibitem{OPSD}
\href{https://data.open-power-system-data.org/time_series/2018-03-13/.}{Open
  {Power} {System} {Data}. 2018. {Data} {Package} {Time} series. {Version}
  2018-03-13. ({Primary} data from various sources, for a complete list see
  {URL}).}
\newline\urlprefix\url{https://data.open-power-system-data.org/time_series/2018-03-13/.}

\bibitem{RESTORE}
A.~Kies, K.~Chattopadhyay, L.~von Bremen, E.~Lorenz, D.~Heinemann, {RESTORE}
  2050 {Work} {Package} {Report} {D12}: {Simulation} of renewable feed-in for
  power system studies., Tech. rep. (2016).

\bibitem{CFSR}
S.~{Saha}, S.~{Moorthi}, X.~{Wu}, J.~{Wang}, S.~{Nadiga}, P.~{Tripp},
  D.~{Behringer}, Y.-T. {Hou}, H.~ya~{Chuang}, M.~{Iredell}, M.~{Ek},
  J.~{Meng}, R.~{Yang}, M.~P. {Mendez}, H.~{van den Dool}, Q.~{Zhang},
  W.~{Wang}, M.~{Chen}, E.~{Becker},
  \href{https://doi.org/10.5065/D6N877VB}{{NCEP} climate forecast system
  version 2 ({CFSv2}) selected hourly time-series products} (2011).
\newline\urlprefix\url{https://doi.org/10.5065/D6N877VB}

\bibitem{Andresen_2015}
G.~B. Andresen, A.~A. S\"ondergaard, M.~Greiner,
  \href{http://www.sciencedirect.com/science/article/pii/S0360544215012815}{Validation
  of {Danish} wind time series from a new global renewable energy atlas for
  energy system analysis}, Energy 93~(Part 1) (2015) 1074 -- 1088.
\newblock \href
  {http://dx.doi.org/https://doi.org/10.1016/j.energy.2015.09.071}
  {\path{doi:https://doi.org/10.1016/j.energy.2015.09.071}}.
\newline\urlprefix\url{http://www.sciencedirect.com/science/article/pii/S0360544215012815}

\bibitem{Victoria_2019b}
M.~Victoria, G.~B. Andresen,
  \href{https://onlinelibrary.wiley.com/doi/full/10.1002/pip.3126}{Using
  validated reanalysis data to investigate the impact of the {PV} system
  configurations at high penetration levels in european countries}, Progress in
  Photovoltaics: Research and Applications 27~(7)  576--592.
\newblock \href {http://dx.doi.org/10.1002/pip.3126}
  {\path{doi:10.1002/pip.3126}}.
\newline\urlprefix\url{https://onlinelibrary.wiley.com/doi/full/10.1002/pip.3126}

\bibitem{Natura2000}
\href{https://www.eea.europa.eu/data-and-maps/data/natura-8}{Natura 2000 data -
  the {European} network of protected sites}.
\newline\urlprefix\url{https://www.eea.europa.eu/data-and-maps/data/natura-8}

\bibitem{Corine_2014}
\href{https://www.eea.europa.eu/data-and-maps/data/clc-2006-raster-4}{Corine
  land cover 2006}, Tech. rep., EEA (2014).
\newline\urlprefix\url{https://www.eea.europa.eu/data-and-maps/data/clc-2006-raster-4}

\bibitem{Quayle_1979}
R.~G. Quayle, H.~F. Diaz,
  \href{https://journals.ametsoc.org/doi/10.1175/1520-0450%281980%29019%3C0241%3AHDDDAT%3E2.0.CO%3B2}{Heating
  {Degree} {Day} {Data} {Applied} to {Residential} {Heating} {Energy}
  {Consumption}}, Journal of Applied Meteorology 19~(3) (1979) 241--246.
\newblock \href
  {http://dx.doi.org/10.1175/1520-0450(1980)019<0241:HDDDAT>2.0.CO;2}
  {\path{doi:10.1175/1520-0450(1980)019<0241:HDDDAT>2.0.CO;2}}.
\newline\urlprefix\url{https://journals.ametsoc.org/doi/10.1175/1520-0450%281980%29019%3C0241%3AHDDDAT%3E2.0.CO%3B2}

\bibitem{NUTS3}
\href{https://data.europa.eu}{Population density by {NUTS} 3 region}.
\newline\urlprefix\url{https://data.europa.eu}

\bibitem{HRE}
\href{www.heatroadmap.eu}{Deliverable 3.1: {Profile} of heating and cooling
  demand in 2015. {Data} {Annex}. {Heat} {Roadmap} {Europe}}.
\newline\urlprefix\url{www.heatroadmap.eu}

\bibitem{BASt}
Verkehrszahlung - {Stundenwerte}, {Tech.} {Rep.}, {Bundesanstalt} f{\"u}r
  {S}tra{\ss}enwesen.

\end{thebibliography}

\appendix
\section{PyPSA-Eur-Sec-30 Model} \label{annex_pypsa_model}

The hourly-resolved, one-node-per-country-network model PyPSA-Eur-Sec-30 is used to optimise the capacity and dispatch in every scenario.\\
 
\begin{figure}[!t]
  \begin{adjustbox}{scale=0.60,trim=5 6.8cm 0 0}
  \begin{circuitikz}
  \draw (1.5,14.5) to [short,i^=grid connection] (1.5,13);
  \draw [ultra thick] (-5,13) node[anchor=south]{electric bus} -- (6,13);
  \draw(2.5,13) |- +(0,0.5) to [short,i^=$$] +(2,0.5);
  \draw (0,-0.5) ;
  \draw (0.5,13) -- +(0,-0.5) node[sground]{};
  \draw (2.5,12) node[vsourcesinshape, rotate=270](V2){}
  (V2.left) -- +(0,0.6);
  \draw (2.5,11.2) node{generators};
    \node[draw,minimum width=1cm,minimum height=0.6cm,anchor=south west] at (3.4,11.9){storage};
    \draw (4,13) to (4,12.5);

  \draw [ultra thick] (-6,10) node[anchor=south]{transport} -- (-3,10);
  \draw (-5.5,10) -- +(0,-0.5) node[sground]{};
  \draw (-3.5,10) to [short,i_=${}$] (-3.5,13);
  \draw (-3.2,11.5)  node[rotate=90]{discharge};
  \draw (-4.5,13) to [short,i^=${}$] (-4.5,10);
  \draw (-4.2,11.5)  node[rotate=90]{charge};
  \node[draw,minimum width=1cm,minimum height=0.6cm,anchor=south west] at (-4.5,8.9){battery};
  \draw (-4,10) to (-4,9.5);

    \draw [ultra thick] (2,10) -- (6.5,10)  node[anchor=south]{heat};
  \draw (3.5,10) -- +(0,-0.5) node[sground]{};
  \draw (4.5,9.35) to [esource] (4.5,8.5);
  \draw (4.5,10) -- (4.5,9.35);
  \draw (4.5,8.3) node{solar thermal};
  \draw (5,13) to [short,i^=heat pump;] (5,10);
  \draw (6.2,11) node{resistive heater};
  \node[draw,minimum width=1cm,minimum height=0.6cm,anchor=south west] at (5.5,8.9){hot water tank};
  \draw (6,10) to (6,9.5);

  \draw [ultra thick] (-2,10)  -- (0.5,10) node[anchor=south]{hydrogen};
  \draw (-1.5,13) to [short,i_=${}$] (-1.5,10);
    \draw (-1.2,11.5)  node[rotate=90]{electrolysis};
  \draw (-0.5,10) to [short,i^=${}$] (-0.5,13);
  \draw (-0.2,11.5)  node[rotate=90]{fuel cell};
  \draw (-1,10) to (-1,9.5);
  \node[draw,minimum width=1cm,minimum height=0.6cm,anchor=south west] at (-1.5,8.9){store};

  \draw [ultra thick] (0.5,8) node[anchor=south]{methane} -- (3,8);
  \draw (1.5,8) to [short,i_=${}$] (1.5,13);
  \draw (2.5,8) to [short,i_=${}$] (2.5,10);
  \draw (1.8,9.2)  node[rotate=90]{generator/CHP};
  \draw (2.8,9)  node[rotate=90]{boiler/CHP};

  \end{circuitikz}

\end{adjustbox}
\caption{Energy flow at a single node representing a country. Within each node, there is a bus (thick horizontal line) for every sector 
(electricity, transport and heating), to which different loads (triangles), energy sources (circles), storage units (rectangles) and converters (lines connecting buses) are attached.}
\label{Fig_buses}
\end{figure}
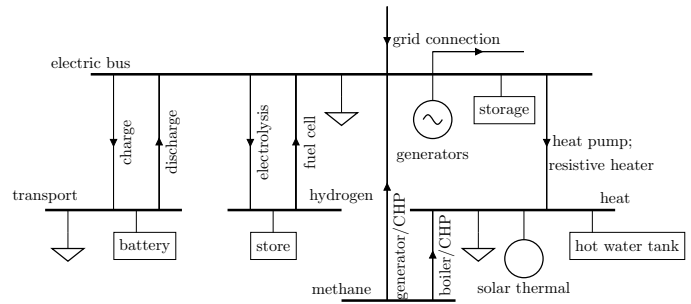

The optimisation objective, that is, the total annualised system cost is calculated as:

\begin{align}
& \min_{\substack{G_{n,s},E_{n,s},\\F_\ell,g_{n,s,t}}} \left[ \sum_{n,s} c_{n,s} \cdot G_{n,s} +\sum_{n,s} \hat{c}_{n,s} \cdot E_{n,s} \right. \nonumber \\
& \hspace{2cm} \left. + \sum_{\ell} c_{\ell} \cdot F_{\ell}+ \sum_{n,s,t} o_{n,s,t} \cdot g_{n,s,t} \right]
\label{eq_objective}
\end{align}

where $c_{n,s}$ are the fixed annualised costs for generator and storage power capacity $G_{n,s}$ of technology $s$ in every bus $n$, $\hat{c}_{n,s}$ are the fixed annualised costs for storage energy capacity $E_{n,s}$, $c_\ell$ are the fixed annualised costs for bus connectors $F_{\ell}$, and $o_{n,s,t}$ are the variable costs (which in some cases include CO$_2$ tax), for generation and storage dispatch $g_{n,s,t}$ in every hour $t$. Bus connectors $\ell$ include transmission lines but also converters between the buses implemented in every country (see Figure \ref{Fig_buses}), for instance, heat pumps that connect the electricity and heating bus. \\

The optimisation of the system is subject to several constraints. First, hourly demand $d_{n,t}$ in every bus $n$ must be supplied by generators in that bus or imported from other buses. $f_{\ell,t}$ represents the energy flow on the link $l$ and $\alpha_{n,\ell,t}$ indicates both the direction and the efficiency of flow on the bus connectors.  $\alpha_{n,\ell,t}$ can be time dependent such as in the case of heat pumps whose conversion efficiency depends on the ambient temperature.

\begin{equation}
\sum_{s} g_{n,s,t}+ \sum_{\ell} \alpha_{n,\ell,t}\cdot f_{\ell,t} = d_{n,t} \hspace{.2cm} \leftrightarrow \hspace{0.2cm} \lambda_{n,t} \hspace{.3cm} \forall\, n,t \label{eq_energy_balance}
\end{equation}
The Lagrange multiplier $\lambda_{n,t}$,  also known as Karun-Kush-Tucker (KKT),  associated with the demand constraint indicates the marginal price of the energy carrier in the bus $n$, \textit{e.g.}, local marginal electricity price in the electricity bus. \\

Second, the maximum power flowing through the links is limited by their maximum physical capacity $F_{\ell}$. For transmission links, $\ubar{f}_{\ell,t}=-1$ and $\bar{f}_{\ell,t}=1$, which allows both import and export between neighbouring countries. For a unidirectional converter \textit{e.g.}, a heat resistor, $\ubar{f}_{\ell,t}=0$ and $\bar{f}_{\ell,t}=1$ since a heat resistor can only convert electricity into heat.

\begin{equation}
\ubar{f}_{\ell,t} \cdot F_{\ell} \leq f_{\ell,t} \leq \bar{f}_{\ell,t} \cdot F_{\ell} \hspace{1cm} \forall\, \ell,t \; . \label{eq_links}
\end{equation}
For interconnecting transmission lines, the lengths $l_{\ell}$ are set by the distance between the geographical mid-points of each country, so that some of the transmission within each country is also reflected in the optimisation. A factor of 25\% is added to the line lengths to account for the fact that transmission lines cannot be placed as the crow flies due to land use restriction. For the transmission lines capacities $F_{\ell}$, a safety margin of 33\% of the installed capacity is used to satisfy n-1 requirements \cite{Brown_2016}. Linear optimal power flow is applied using Kirchhoff's formulation \cite{Horsch_2018}. \\

Third, for every hour the maximum capacity that can provide a generator or storage is bounded by the product between installed capacity $G_{n,s}$ and availabilities $\ubar{g}_{n,s,t}$, $\bar{g}_{n,s,t}$. For instance, for solar generators $\ubar{g}_{n,s,t}$ is zero and $\bar{g}_{n,s,t}$ refers to the capacity factor at time $t$ 
\begin{equation}
\ubar{g}_{n,s,t} \cdot G_{n,s} \leq g_{n,s,t} \leq \bar{g}_{n,s,t} \cdot G_{n,s} \hspace{1cm} \forall\, n,s,t \; . \label{eq_g}
\end{equation}
The maximum power capacity for generators is limited by potentials $\bar{G}_{n,s}$ that are estimated taking into account physical and environmental constraints:
\begin{equation}\label{eq_max_G}
0 \leq G_{n,s}\leq \bar{G}_{n,s} \hspace{1cm} \forall\, n,s \; .
\end{equation}
The storage technologies have a charging efficiency $\eta_{in}$ and rate $g_{n,s,t}^+$, a discharging efficiency $\eta_{out}$ and rate $g_{n,s,t}^-$, possible inflow $g_{n,s,t,\textrm{inflow}}$ and spillage $g_{n,s,t,\textrm{spillage}}$, and standing loss $\eta_0$. The state of charge $e_{n,s,t}$ of every storage has to be consistent with charging and discharging in every hour and is limited by the energy capacity of the storage $E_{n,s}$. It should be remarked that the storage energy capacity $E_{n,s}$ can be optimised independently of the storage power capacity $G_{n,s}$.
\begin{align}
e_{n,s,t} = & \ \eta_0 \cdot e_{n,s,t-1} + \eta_{in} |g_{n,s,t}^+| - \eta_{out}^{-1} |g_{n,s,t}^-| \nonumber \\
& + g_{n,s,t,\textrm{inflow}} - g_{n,s,t,\textrm{spillage}} \; , \nonumber \\
& 0  \leq   e_{n,s,t} \leq E_{n,s}   \hspace{0.5cm} \forall\, n,s,t \; . \label{eq_storage}
\end{align}
So far, equations (\ref{eq_energy_balance}) to (\ref{eq_storage}) represent mainly technical constraints but additional constraints can be imposed to bound the solution.\\

The interconnecting transmission expansion can be limited by a global constraint
\begin{equation}
\sum_{\ell} l_\ell \cdot F_{\ell} \leq  \textrm{CAP}_{LV} \hspace{.7cm} \leftrightarrow \hspace{0.3cm} \mu_{LV} \; ,
\label{eq_cap}
\end{equation}
where the sum of transmission capacities $F_{\ell}$ multiplied by the lengths $l_{\ell}$ is bounded by a transmission volume cap $\textrm{CAP}_{LV}$. In this case, the Lagrange/KKT multiplier $\mu_{LV}$ represents the shadow price of a marginal increase in transmission volume.\\

The average gross VRES generation $g_{i,VRES}^{gross}$ is imposed to be proportional to the average electricity demand $d_{elec,i,t}$ in every country through the constraint
 
\begin{equation} \label{eq_gamma}
< g_{i,VRES}^{gross} > = \gamma_i < d_{elec,i,t} >, \hspace{0.4cm} \gamma_i = \gamma
\end{equation}
The adjective `gross' indicates that $\gamma_i$ is calculated based on the energy that can be potentially generated, that is before curtailment, and including onshore and offshore wind as well as solar photovoltaics. This `weakly-homogeneous' constraint was proposed in \cite{Eriksen_2017} and used in \cite{Zhu_2019}, but it should be mentioned that it was not included in the implementation in \cite{Brown_2018, Brown_2019}.\\

Finally, the maximum CO$_2$ allowed to be emitted by the system $\textrm{CAP}_{CO2}$ can be imposed through the constraint 

\begin{equation}
  \sum_{n,s,t}  \varepsilon_{s} \frac{ g_{n,s,t} }{\eta_{n,s}}  \leq  \textrm{CAP}_{CO2} \hspace{.4cm} \leftrightarrow \hspace{0.3cm} \mu_{CO2} \label{eq_co2cap}
\end{equation}
where $\varepsilon_{s}$ represents the specific emissions in CO$_2$-tonne-per-MWh\th{} of the fuel $s$, $\eta_{n,s}$ the efficiency and $g_{n,s,t}$ the generators dispatch. In this case, the Lagrange/KKT multiplier represents the shadow price of CO$_2$, \textit{i.e.}, the additional price that should be added for every unit of CO$_2$ to achieve the CO$_2$ reduction target in an open market. 

\section{Data} \label{annex_data} \label{anex_data}

\paragraph{Electricity sector}\ \label{sec_electricity}

The hourly electricity demand in every country is represented by means of historical data from 2015 reported by the European Network of Transmission System Operators for Electricity (ENTSO-e) \cite{ENTSOE} through the convenient data collection provided by the Open Power System Data (OPSD) initiative \cite{OPSD}. The electricity used to supply domestic space and water heating demand is subtracted from historical data since it is included in the heating demand time series (described below). The total annual electricity demand accounts for 2,854 TWh and its profile is shown in Figure \ref{fig_demands} in the main text. Electricity can be generated by means of onshore and offshore wind, solar photovoltaic and Open Cycle Gas Turbines (OCGT) whose capacities are optimised, as well as by run-of-river and reservoir-based hydroelectric power plants. Hydroelectric energy and power capacities are fixed exogenously according to current values \cite{RESTORE}. Table \ref{tab_costs} gathers the costs, lifetime, and efficiency values assumed for the different technologies. \\

The availability of renewable generators is represented by means of the hourly capacity factors that have been calculated using the Climate Forecast System Reanalysis (CFSR) dataset \cite{CFSR} and the Global Renewable Energy Atlas (REatlas) \cite{Andresen_2015, Victoria_2019b}. For solar PV, reanalysis irradiance is first bias-corrected based on satellite data and then used to generate time series that are validated with historical data \cite{Victoria_2019b}. The dataset for solar capacity factor time series is openly available and can be retrieved from \href{https://doi.org/10.5281/zenodo.1321809}{10.5281/zenodo.1321809}. For onshore and offshore wind capacity factor time series, a capacity layout in which the installed capacity is proportional to the wind resource in every CFSR raster cell is assumed. Following \cite{Schlachtberger_2017}, the onshore wind layout of the ten largest countries is split into up to four equal-area parts. Independent classes of generators with different time series and average full load hours are added to a single node representing a country. Their optimised capacities are later aggregated on a country level for analysis. To calculate the maximum capacities that can be installed in every raster cell the protected sites listed in Natura 2000 \cite{Natura2000} and non-adequate land types, \textit{e.g.} cities as specified by \cite{Corine_2014}, are excluded. For the remaining areas, a maximum 20\% and 1\% coverage ratio is used for wind and solar PV respectively which results in a maximum density of 2 MW/km$^2$ for wind and 1.5 MW/km$^2$ for PV. For offshore wind, the maximum water depth assumed is 50 m. For every country, the renewable capacity is extended until the installation density reaches its maximum in the exploitable area of any of the CFSR rater cells.  \\

Regarding PV, the model assumes that 50\% of the installed capacities correspond to rooftop mounted systems while the other 50\% belongs to larger utility-scale power plants with different costs and discount rates assumed for every type of installation. 

\paragraph{Heating sector}\

Hourly heating demand time series are obtained for 2015 by using the Heating Degree Hour (HDH) approximation \cite{Quayle_1979} and hourly ambient temperature values from the CFSR reanalysis dataset. The HDH approach assumes that the heating demand increases linearly from a threshold temperature of 17$^{\circ}$C. HDH time series at every grid point in the CFSR are weighted by population density (the NUTS3 population data \cite{NUTS3} is used as a proxy) and aggregated at a national scale to obtain HDH time series representative of a country. The time series are then scaled based on the annual demands for domestic space heating in 2015, which are retrieved from the Heat Roadmap Europe project \cite{HRE}. Finally, a constant hourly value for the hot water consumption, obtained from the same database, is added to compute the total heating demand time series representative for every country. The estimated values for total annual demand in Europe are similar for electricity and heating, accounting for 2,854 TWh\el/a and 3,562 TWh\th/a respectively but heating demand shows a much more pronounced seasonal variation, see Figure \ref{fig_demands}.\\

The technologies available in the model that supply heat include gas boilers, resistive heaters, heat pumps and CHP units. A lower cost can be achieved if those conversion technologies are built at a larger scale to feed-in district heating systems. Since centralised solutions are only cost-effective when the population density is above a certain threshold, the heating bus is split in two in the model (more details can be found in \cite{Brown_2018, Zhu_2019}), the overall heating demand is divided proportionally into \textit{urban heating} and \textit{rural heating}. The \textit{urban heating} bus supplies heating demand in places whose population density is high enough to allow district heating. The cost-optimal solution can include district heating for every country in Europe except for southern countries (Spain, Greece, Portugal, Italy, and Bulgaria) where the high winter temperatures reduce the competitiveness of district heating systems. Where district heating is allowed, CHP plants can be built to feed-in the system. Central ground-sourced heat pumps have been assumed in urban areas with district heating, while air heat-sourced pumps have been assumed for individual systems in urban areas. On the other hand, the \textit{rural heating} bus represents the places where only decentralised heating units are allowed. In this case, individual ground-sourced heat pumps have been assumed. Despite being more expensive, the higher COP of ground-sourced heat pumps makes them economically favourable. Additionally, in both heating buses, resistive heaters or gas boilers can also be used to supply the demand. The description of the models for the temperature-dependent efficiency of heat pumps and possible electricity-heat output combinations for CHP, as well as the efficiencies assumed for gas boilers and resistive heaters, are included in \cite{Brown_2018} and summarised in Table \ref{tab_costs}. Long-term central thermal energy storage (CTES), \textit{i.e.} well-insulated hot water tanks in pits containing tens of thousands of cubic metres of hot water, can be built in urban heating buses where district heating is allowed and individual thermal energy storage (ITES), \textit{i.e.} small water tanks, can be built where only individual solutions are allowed.

\paragraph{Transport sector} \

Annual energy demands from road and rail transport for every country are retrieved from \cite{ODYSSEE}. Aviation, shipping, and pipe transport are not included in the model. Road and rail transport are considered to be fully electrified, a country-specific factor (averaging 3.5) is used to account for the increased efficiency when electrifying transport. Country-specific factors are computed by comparing the current car final energy consumption per km in \cite{ODYSSEE} (averaging 0.7 kWh/km) to the 0.2 kWh/km value assumed for plug-to-wheels efficiency in EVs. The characteristic weakly profile provided by the German Federal Highway Research Institute (BASt) \cite{BASt} is used to obtain hourly time series for European countries taking into account the corresponding local times. Furthermore, a temperature dependence is included in the time series to account for heating/cooling demand in transport. For temperatures below/above 15$^{\circ}$C/20$^{\circ}$C, temperature coefficients of 0.63\%/$^{\circ}$C and 0.98\%/$^{\circ}$C are assumed, see \cite{Brown_2018} for more details. The annual electricity demand from transportation sector in Europe accounts for 1,102 TWh/a and the profile is shown in Figure \ref{fig_demands}.  

\paragraph{Costs and CO$_2$ emissions} \

Most of the costs assumed in the model are based on predictions for 2030. They are annualised assuming a discount rate of 7\%. Table \ref{tab_costs} and \ref{tab_stores} summarise the investment and operation and maintenance costs, lifetimes, and efficiency values assumed in the model. The CO$_2$ emissions in the model come only from those technologies using gas (OCGT, CHP, and gas boilers). An emission factor equal to 0.19 tCO$_2$/MWh$_{th}$ and a cost of 21.6 \EUR/MWh$_{th}$ have been assumed for gas. Coal and nuclear are not considered in this paper. The three sectors included in the model emitted 3016 megatonnes of CO$_2$ (MtCO$_2$) in 1990 \cite{ODYSSEE}.  Electricity generation was responsible for 1510 MtCO$_2$, land-based transport for 784 MtCO$_2$, and heating in the residential and service sectors for 723 MtCO$_2$. The emission reduction of 95\% compared to 1990 level assumed in Section \ref{sec_sensitivity} for the Electricity sector corresponds to a limit of 76 MtCO$_2$ per year.

\paragraph{Code} \

The model used in this paper is the PyPSA-Eur-Sec-30 whose code was used in \cite{Brown_2018, Brown_2019, Zhu_2019} and can be accessed through the repository \href{https://doi.org/10.5281/zenodo.1146666}{10.5281/zenodo.1146666}. Moreover, additional code to plot the figures is available at \href{https://github.com/martavp/PyPSA-plots.git}{github.com/martavp/PyPSA-plots.git}.

\end{document}